\documentclass[journal]{IEEEtran}

\usepackage[T1]{fontenc}
\usepackage{lmodern}
\usepackage{booktabs}
\usepackage{graphicx}
\usepackage{float}
\usepackage{listings}
\usepackage{xcolor}
\usepackage{xspace}
\usepackage{enumitem}
\usepackage{amsmath}
\usepackage{mdframed}
\usepackage{microtype}
\usepackage{hyperref}
\hypersetup{colorlinks=true, urlcolor=blue, linkcolor=blue, citecolor=blue}
\usepackage{xurl}
\usepackage{cleveref}
\crefname{figure}{Figure}{Figures}
\Crefname{figure}{Figure}{Figures}
\crefname{table}{Table}{Tables}
\Crefname{table}{Table}{Tables}

\usepackage[most]{tcolorbox}

\usepackage{algorithm}
\usepackage{algorithmic}

\usepackage{tikz}
\usetikzlibrary{arrows.meta,shapes,positioning,shapes.geometric,fit,backgrounds,calc}

\newcommand{\sysname}{\textsc{AssumptionMiner}\xspace}

\newcommand{\Aone}{$\mathsf{A_1}$\xspace}
\newcommand{\Atwo}{$\mathsf{A_2}$\xspace}
\newcommand{\Athree}{$\mathsf{A_3}$\xspace}
\newcommand{\Afour}{$\mathsf{A_4}$\xspace}
\newcommand{\Afive}{$\mathsf{A_5}$\xspace}

\newcommand{\Bone}{$\mathsf{B_1}$\xspace}
\newcommand{\Btwo}{$\mathsf{B_2}$\xspace}
\newcommand{\Bthree}{$\mathsf{B_3}$\xspace}
\newcommand{\Bfour}{$\mathsf{B_4}$\xspace}

\newtcolorbox{rqbox}[1]{
  colback  = gray!8,
  colframe = black!55,
  fonttitle= \bfseries\small,
  title    = {#1},
  left     = 4pt,
  right    = 4pt,
  top      = 3pt,
  bottom   = 3pt,
  before upper = {\small},
}

\newcommand{\code}[1]{\texttt{#1}}

\newcommand{\AR}{$\mathit{AR}$\xspace}
\newcommand{\AP}{$\mathit{AP}$\xspace}
\newcommand{\AFone}{$\mathit{AF_1}$\xspace}

\begin{document}

\title{AssumptionMiner: Extracting, Tracing, and Revising Implicit Assumptions in LLM Code Generation}

\author{Jie JW Wu%
\IEEEcompsocitemizethanks{
  \IEEEcompsocthanksitem Jie JW Wu is with Michigan Technological University,
  Houghton, MI, USA.
  E-mail: jie.jw.wu@mtu.edu. ORCID: 0000-0002-7895-2023.
}
}

\IEEEtitleabstractindextext{%
\begin{abstract}
Large language models (LLMs) generate code from natural-language prompts, but
real prompts are rarely complete specifications.
When a prompt leaves input formats, error handling, or design decisions
unstated, the model fills the gaps with \emph{implicit assumptions}: choices
absent from the prompt that nonetheless shape the code's behavior, and often
its correctness.
Because these assumptions are invisible, generated code can pass its tests
while violating the developer's intent.

We present \sysname, a framework that makes such assumptions a first-class
output of code generation.
Instead of producing code alone, \sysname also emits an explicit
\emph{assumption layer}: a structured record of every inferred constraint and
design decision.
Developers inspect, confirm, or revise assumptions through an interactive
interface, and an AST-based dependency graph lets the system regenerate only
the code a revised assumption governs.

We also contribute a benchmark of 180 ambiguous programming tasks with 676
annotated assumptions, including a human-verified curated subset for
code-localization evaluation.
We evaluate extraction, code localization, and revision support.
With open-source models, a confidence-weighted ensemble attains $F_1=0.816$
for extraction, improving on the strongest offline baseline by $3.6\times$.
Under a stricter decision-level metric that also requires each extracted
description to name the same design decision as the reference, the best
open-source configuration reaches only $F_1=0.66$, far above the near-zero
non-LLM baselines but low in absolute terms, indicating that decision-level
assumption extraction remains far from solved.
On the human-verified localization set, AST-guided localization produces
tighter code regions than keyword and whole-file baselines. When developers revise an assumption, targeted
regeneration changes less code than non-targeted alternatives, although it
also exposes a need for stronger handling of cascading edits.
Together, these results show how making assumptions explicit can support more
inspectable and controllable LLM-based code generation.
The benchmark, code, and replication package are available at
\url{https://doi.org/10.5281/zenodo.21535058}.
\end{abstract}

}

\maketitle
\IEEEdisplaynontitleabstractindextext
\IEEEpeerreviewmaketitle

\section{Introduction}
\label{sec:intro}

\begin{figure*}[t]
  \centering
  \resizebox{\textwidth}{!}{%
  \begin{tikzpicture}[
    cl/.style={anchor=west, font=\ttfamily\small, inner sep=2pt,
               minimum height=0.40cm, minimum width=8.0cm},
    rcl/.style={anchor=west, font=\ttfamily\small, inner sep=2pt,
                minimum height=0.40cm, minimum width=3.8cm},
    badge/.style={draw, rounded corners=1.5pt, font=\scriptsize\bfseries,
                  text=white, inner sep=2pt, minimum width=0.55cm,
                  minimum height=0.26cm, anchor=center},
    acard/.style={draw=gray!55, rounded corners=2pt, thick, fill=white,
                  font=\small, inner sep=3pt, anchor=west,
                  text width=5.0cm, minimum height=0.42cm},
    aedit/.style={draw=green!55!black, rounded corners=2pt, ultra thick,
                  fill=green!5, font=\small, inner sep=3pt, anchor=west,
                  text width=5.0cm, minimum height=0.42cm},
    arr/.style={-{Stealth[scale=0.5]}, thin, dashed, gray!55},
    earr/.style={-{Stealth[scale=0.5]}, green!55!black, semithick},
  ]
  \draw[draw=gray!40, rounded corners=5pt, thick, fill=gray!4]
    (-0.15, 0.80) rectangle (8.4, -3.60);
  \node[font=\small\bfseries] at (4.1, 0.62) {Without \textsc{AssumptionMiner}};
  \node[cl, fill=white]     at (0.1, 0.25) {\texttt{def authenticate(user, pwd):}};
  \node[cl, fill=red!12]    (L1) at (0.1,-0.20) {\texttt{\phantom{xx}h = md5(pwd.encode())}};
  \node[cl, fill=gray!8]    at (0.1,-0.63) {\texttt{\phantom{xx}if h != user.hash: raise}};
  \node[cl, fill=cyan!10]   (L3) at (0.1,-1.06) {\texttt{\phantom{xx}tok = str(uuid.uuid4())}};
  \node[cl, fill=orange!12] (L4) at (0.1,-1.49) {\texttt{\phantom{xx}if fail\_cnt > 3: raise}};
  \node[cl, fill=blue!10]   (L5) at (0.1,-1.92) {\texttt{\phantom{xx}\_sessions[uid] = tok}};
  \node[cl, fill=purple!10] (L6) at (0.1,-2.35) {\texttt{\phantom{xx}for \_ in range(100): ...}};
  \node[cl, fill=white]     at (0.1,-2.78) {\texttt{\phantom{xx}return \{"token": tok\}}};
  \node[badge, fill=red!65]           at (8.00,-0.20) {T6};
  \node[badge, fill=cyan!55!black]    at (8.00,-1.06) {T2};
  \node[badge, fill=orange!70!black]  at (8.00,-1.49) {T3};
  \node[badge, fill=blue!65]          at (8.00,-1.92) {T4};
  \node[badge, fill=purple!65]        at (8.00,-2.35) {T5};
  \node[font=\small\itshape, text=red!65!black, text width=7.8cm, align=center]
    at (4.1, -3.25) {5 implicit assumptions (colored), committed silently,
                     invisible to developer until code review};
  \draw[-{Stealth[scale=1.0]}, very thick, gray!55]
    (8.7, -1.4) -- (9.8, -1.4)
    node[midway, above, font=\scriptsize\bfseries, text=gray!60] {\textsc{AM}};
  \draw[draw=gray!40, rounded corners=5pt, thick, fill=gray!4]
    (10.1, 0.80) rectangle (21.2, -3.60);
  \node[font=\small\bfseries] at (15.65, 0.62) {With \textsc{AssumptionMiner}};
  \node[rcl, fill=white]     at (10.9, 0.25) {\texttt{def authenticate(...):}};
  \node[rcl, fill=red!12]    (R1) at (10.9,-0.20) {\texttt{\phantom{xx}h = md5(pwd)}};
  \node[rcl, fill=gray!8]    at (10.9,-0.63) {\texttt{\phantom{xx}if h != ...: ...}};
  \node[rcl, fill=cyan!10]   (R3) at (10.9,-1.06) {\texttt{\phantom{xx}tok = uuid4()}};
  \node[rcl, fill=orange!12] (R4) at (10.9,-1.49) {\texttt{\phantom{xx}if cnt > 3: ...}};
  \node[rcl, fill=blue!10]   (R5) at (10.9,-1.92) {\texttt{\phantom{xx}sessions[u] = tok}};
  \node[rcl, fill=purple!10] (R6) at (10.9,-2.35) {\texttt{\phantom{xx}for \_ in ...: ...}};
  \node[rcl, fill=white]     at (10.9,-2.78) {\texttt{\phantom{xx}return tok}};
  \node[acard, fill=gray!5]   (C1) at (15.2, 0.05) {\textbf{[T2]}\ A2: token = plain UUID};
  \node[aedit]                (C2) at (15.2,-0.45) {\textbf{[T6]}\ A1: MD5\,$\to$\,\textbf{bcrypt} {\footnotesize(edited)}};
  \node[acard]                (C3) at (15.2,-0.95) {\textbf{[T3]}\ A3: retry limit\,=\,3};
  \node[acard, fill=blue!4]   (C4) at (15.2,-1.45) {\textbf{[T4]}\ A4: in-memory only};
  \node[acard, fill=purple!4] (C5) at (15.2,-1.95) {\textbf{[T5]}\ A5: O(n) linear scan};
  \draw[arr]  (R3.east) -- (C1.west);
  \draw[earr] (R1.east) -- (C2.west);
  \draw[arr]  (R4.east) -- (C3.west);
  \draw[arr]  (R5.east) -- (C4.west);
  \draw[arr]  (R6.east) -- (C5.west);
  \node[font=\scriptsize\itshape, text=green!50!black, text width=5.8cm, align=center]
    at (15.65, -3.25) {Developer edits A1 (MD5 $\to$ bcrypt);\\
                      \textsc{AM} regenerates only the affected region};
  \end{tikzpicture}}%
  \caption{Implicit assumptions in LLM code generation, before and after \sysname.
           \textbf{Left}: An ambiguous prompt yields code with five undisclosed
           assumptions (colored, tagged with taxonomy categories) about password
           storage (A1), session format (A2), failure policy (A3),
           persistence (A4), and performance (A5), invisible until code review.
           \textbf{Right}: \sysname augments the same generation step to jointly
           produce code \emph{and} an explicit assumption layer (A1--A5).
           The developer confirms four assumptions and flips A1 (MD5 $\to$ bcrypt);
           \sysname incrementally regenerates only the affected code region.}
  \label{fig:teaser}
\end{figure*}
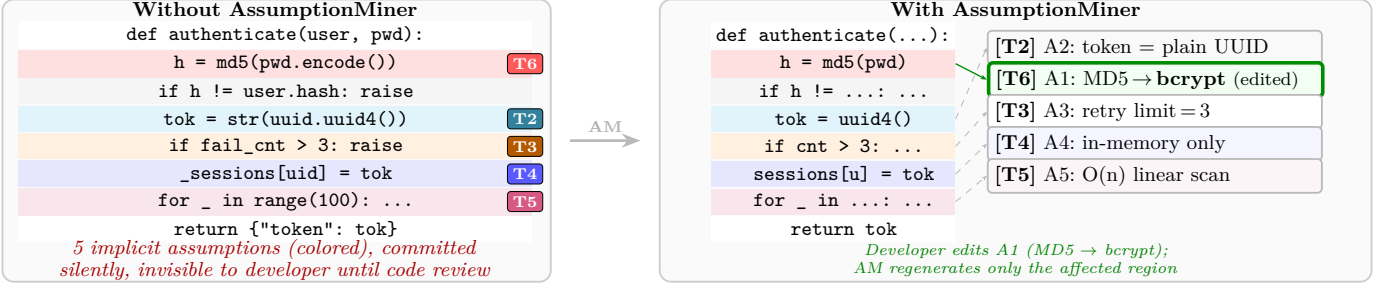

\IEEEPARstart{M}{odern} software development increasingly relies on large language models (LLMs) to
accelerate code generation~\cite{chen2021codex,austin2021mbpp,li2022alphacode,jimenez2024swebench}.
Developers write natural-language prompts and receive executable code within
seconds, a workflow that has moved from research prototypes into everyday
practice~\cite{ziegler2022productivity,vaithilingam2022expectation}.
Yet real-world prompts are rarely complete specifications: they leave open questions about
input formats, boundary conditions, error-handling policies, domain constraints, and
intended semantics~\cite{wiegers2013software,ferrari2018nlrequire}.

This gap is structural, not incidental.
Traditional software engineering moves from requirements to design to
implementation through successive refinements, and each translation step is an
opportunity to surface and resolve ambiguity.
LLM-based code generation collapses these steps: natural language is converted
directly into executable code in a single opaque step.
As Mary Shaw observes, ``vibe coding ignores the shifts from the language of
the world to the language of the machine''~\cite{shaw2026vibe}: the
disambiguation work that those shifts used to force into the open does not
disappear. The model performs it silently.
When an LLM fills a specification gap, it commits to \emph{implicit
assumptions}: specific design choices that were never stated in the prompt and
are never reported to the developer.

These implicit assumptions are a pervasive source of misalignment.
The generated code may pass unit tests while violating project conventions, security
policies, or the developer's actual intent, a class of defect we term \emph{working-but-wrong}.
In our benchmark of 180 LLM-generated code artifacts, every task contained at
least two latent implicit assumptions (mean: 3.8 per task, range: 2--5) identifiable
only through inspection, motivating the need for automated extraction.

The core difficulty is that existing tools give developers \emph{no visibility}
into what the LLM assumed.
Prompt engineering guides~\cite{white2023prompting} advise more detailed prompts but
cannot enumerate what was left out.
Stoica et al.\ observe that when a request is ambiguous, an LLM-based agent
has two simple ways to disambiguate: it can \emph{ask the user for
clarification}, or it can \emph{state the assumptions} underlying its answer
so that the user sees how to improve the
prompt~\cite{stoica2024specifications}.
The first strategy has grown into an active line of clarification-question
research: ClarifyGPT detects ambiguous requirements via code-consistency
checks and asks targeted questions~\cite{mu2024clarifygpt}; SpecValidator instead
detects defective task descriptions (lexical vagueness, under-specification, and
malformed formatting) with a lightweight fine-tuned
classifier~\cite{akli2026defective}; HumanEvalComm
benchmarks whether code LLMs ask clarifying questions at all when requirements
are inconsistent, ambiguous, or incomplete, finding that they usually generate
code instead~\cite{wu2025humanevalcomm}, and such defective descriptions
measurably degrade code-model accuracy~\cite{larbi2025prompts}; ClarifyCoder
fine-tunes models to prefer asking over answering~\cite{wu2025clarifycoder};
TiCoder interactively formalizes user intent by generating tests for the user
to confirm~\cite{lahiri2022interactive}; and self-planning approaches decompose
prompts before generation~\cite{jiang2023selfplanning}.
These systems reduce ambiguity up-front, but their questions are hypothetical
(what \emph{might} matter) rather than diagnostic (what the model
\emph{actually decided}), their answers are not linked to specific code
regions, and a later change of decision still requires a full reprompt.
The second strategy, making the model's assumptions themselves explicit, has
remained largely unexplored for code generation, even though it imposes no
up-front interaction cost on the developer.
Self-repair methods~\cite{olausson2023selfrepair,zheng2023selfrefine} iterate on test
failures but cannot repair silent assumptions that produce passing-but-wrong code.

In this research, we operationalize the second strategy: we present \sysname, a framework
that makes implicit assumptions a \emph{first-class output} of LLM code
generation.
Rather than extracting assumptions after the fact, \sysname augments the generation step
itself: the backbone LLM jointly produces executable code \emph{and} an explicit
\emph{assumption layer}: a structured, formalized list of every inferred constraint and
design decision embedded in the generated code.
Where clarification-question systems require the developer (or model) to know
in advance what to ask, \sysname \emph{diagnoses} which decisions the LLM
already made, surfacing them as revisable records grounded in the actual
generated code.
Developers inspect, confirm, or modify individual assumptions through an interactive
interface (\S\ref{sec:interface}). When an assumption is flipped or revised, \sysname
incrementally regenerates only the code regions governed by that assumption while
preserving all unaffected components.

\noindent\textbf{Contributions.} We make the following contributions in this research:
\begin{enumerate}[leftmargin=*, label=(C\arabic*)]
  \item \textbf{Assumption taxonomy} (\S\ref{sec:taxonomy}): a six-category
        taxonomy of implicit assumptions in LLM-generated code, covering input
        validation, data format, error policy, persistence, performance, and
        security.
  \item \textbf{Implicit-assumption generation framework} (\S\ref{sec:framework}): a novel
        approach that integrates assumption surfacing directly into LLM-based code
        synthesis, producing code and a formalized assumption layer in a single augmented
        generation step, making implicit LLM reasoning explicit and inspectable.
  \item \textbf{Assumption--code dependency mechanism} (\S\ref{sec:dependency},
        \S\ref{sec:regeneration}): a two-pass AST-guided dependency graph (keyword-overlap
        candidate identification followed by minimal-subtree AST narrowing), linking each
        assumption to the code regions it governs, enabling traceable and controllable
        incremental code updates when assumptions are flipped or revised.
  \item \textbf{Interactive interface and software artifact} (\S\ref{sec:interface},
        Appendix~\ref{app:software}): a developer-facing review interface and an
        open-source implementation of the full pipeline.
  \item \textbf{Benchmark and empirical evaluation}
        (\S\ref{sec:benchmark}, \S\ref{sec:evaluation}): 180 programming tasks
        under ambiguous prompts, with 676 annotated assumptions (including a
        30-task curated subset of novel tasks with 109 human-verified assumptions), evaluated
        across extraction quality, dependency accuracy, and code adaptability.
\end{enumerate}

\noindent\textbf{Paper organization.}
\S\ref{sec:background} defines implicit assumptions and motivates the problem with a
running example.  \S\ref{sec:framework} describes the \sysname architecture.
\S\ref{sec:interface} presents the developer interface.  \S\ref{sec:benchmark} introduces
the benchmark.  \S\ref{sec:evaluation} reports experimental results and discussion.
\S\ref{sec:related} surveys related work.  \S\ref{sec:threats} discusses threats to
validity.  \S\ref{sec:conclusion} concludes.
Appendices provide the full extraction prompt (\S\ref{app:prompt}), complete per-category
P/R/$F_1$ tables (\S\ref{app:tables}), benchmark construction details
(\S\ref{app:benchmark}), the ensemble algorithm (\S\ref{app:ensemble}),
software-artifact details (\S\ref{app:software}), and
reproducibility notes (\S\ref{app:repro}).
Our replication package (the benchmark, source code, all evaluation scripts,
and raw results) is publicly available at
\url{https://doi.org/10.5281/zenodo.21535058}.

\section{Motivation \& Background}
\label{sec:background}

\subsection{Formal Definition}
\label{sec:definition}

Let $p$ be a natural-language prompt (potentially ambiguous or underspecified) and
$c = \mathcal{G}(p)$ be the code produced by an LLM $\mathcal{G}$.
We define an \emph{implicit assumption} as follows.

\begin{mdframed}[linewidth=0.8pt, innerleftmargin=6pt, innerrightmargin=6pt,
                 innertopmargin=4pt, innerbottommargin=4pt]
\textbf{Definition 1 (Implicit Assumption).}
A \emph{implicit assumption} $a$ of $(p, c)$ is a proposition that:
\begin{enumerate}[label=(\roman*), noitemsep, topsep=2pt]
  \item is not stated in $p$ (arising from ambiguity or underspecification),
  \item is instantiated in $c$ (i.e., $c$ behaves as if $a$ holds), and
  \item has at least one alternative instantiation that satisfies $p$ but would
        produce a meaningfully different $c$.\footnote{We operationalize
        ``meaningfully different'' as producing distinct observable behavior on
        at least one input not excluded by~$p$. For example, using MD5 vs.\
        bcrypt for password hashing yields different outputs for any password
        string, satisfying this condition.}
\end{enumerate}
\end{mdframed}

\noindent The set of implicit assumptions of $(p, c)$ is denoted $\mathcal{A}(p,c)$.
In \sysname, $\mathcal{A}(p,c)$ is produced \emph{during} generation as a formalized
\emph{assumption layer} co-output alongside $c$.
Each assumption $a_i \in \mathcal{A}(p,c)$ is characterized by:
(i) its \emph{category} $\kappa(a_i)$ (see \S\ref{sec:taxonomy}),
(ii) a natural-language \emph{description} $d(a_i)$,
(iii) the \emph{code regions} $R(a_i) \subseteq c$ it governs, and
(iv) a set of \emph{alternatives} $\text{Alt}(a_i)$ representing choices the LLM
     could have made differently.

\subsection{Running Example}
\label{sec:example}

We use the following prompt as a running example throughout the paper:

\begin{mdframed}[linewidth=0.8pt, backgroundcolor=gray!7,
                 innerleftmargin=8pt, innerrightmargin=8pt,
                 innertopmargin=4pt, innerbottommargin=4pt]
\emph{``Implement a user authentication system that checks credentials
and returns a session token.''}
\end{mdframed}

A typical LLM-generated implementation embeds at least five implicit assumptions,
which \sysname labels \Aone--\Afive:

\begin{description}[leftmargin=2em, labelindent=0pt, noitemsep]
  \item[\Aone \textbf{Password storage (T6)}:] Passwords are hashed with MD5
       (weak, SHA-256 or bcrypt preferred).
  \item[\Atwo \textbf{Session format (T2)}:] Tokens are plain UUID v4 strings, with no signing or expiry logic.
  \item[\Athree \textbf{Failure policy (T3)}:] After three failed attempts a bare
       exception is raised, with no structured error or HTTP status code.
  \item[\Afour \textbf{Persistence (T4)}:] Sessions are held in an in-memory
       Python dict, with no database or file persistence.
  \item[\Afive \textbf{Performance (T5)}:] A fixed-iteration loop performs an
       $O(n)$ linear scan, with no indexing or configurable work factor.
\end{description}

These five assumptions are silently embedded in fewer than a dozen lines of
generated code (see \cref{fig:teaser}).  Without a tool like \sysname, the
developer would need to audit the generated code manually to discover them.
(Category labels T1--T6 refer to the assumption taxonomy introduced in
\S\ref{sec:taxonomy}.)

\paragraph{Second Example: An Algorithmic Function.}
To demonstrate that implicit assumptions pervade even short algorithmic code,
consider a prompt typical of competitive programming or coding interviews:

\begin{mdframed}[linewidth=0.8pt, backgroundcolor=gray!7,
                 innerleftmargin=8pt, innerrightmargin=8pt,
                 innertopmargin=4pt, innerbottommargin=4pt]
\emph{``Given a list of integers and a target value, find two numbers that add up to the target.''}
\end{mdframed}

A hash-map solution generated by GPT-4o in seven lines embeds four implicit
assumptions invisible in the prompt:

\begin{description}[leftmargin=2em, labelindent=0pt, noitemsep]
  \item[\Bone \textbf{Return type (T2):}] The function returns a pair of
       \emph{indices} (e.g., \code{[1,~3]}), not the \emph{values}
       themselves, a choice the prompt does not specify.
  \item[\Btwo \textbf{Uniqueness (T3):}] Exactly one valid pair is assumed. The first match is returned without checking for additional solutions.
  \item[\Bthree \textbf{No-solution policy (T3):}] An empty list \code{[]}
       is returned when no pair sums to the target, not \code{None}, not a
       raised \code{ValueError}.
  \item[\Bfour \textbf{Space--time trade-off (T5):}] An $O(n)$-space hash map
       is chosen over an $O(n^2)$ brute-force scan or an $O(n\log n)$
       sort-and-scan approach.
\end{description}

Unlike the authentication example, these assumptions arise in seven lines of
purely algorithmic code with no external dependencies, confirming that implicit
assumptions are not confined to system-design tasks: they emerge whenever a
prompt leaves behavioral or design questions open, regardless of code complexity.

\subsection{Assumption Taxonomy}
\label{sec:taxonomy}

We developed the taxonomy through an iterative, author-led open-coding process
over 30 LLM-generated code samples from the curated subset of the benchmark,
assisted by LLM-proposed candidate labels~\cite{strauss1998basics}.
In each iteration, candidate assumptions and category labels were first proposed
automatically (GPT-4o), then reviewed, corrected, and grouped into emergent
themes by the author. Category boundaries were refined between iterations.
The most contested boundary during this process was T1/T3: a guard clause is T1
(Input Validation) when it enforces a caller precondition, but T3 (Error Policy)
when the design question is \emph{how} the function signals failure, a
distinction that matters for code reuse across call sites.
The process converged after three iterations. The same six categories covered
the tasks in the full benchmark without requiring additional categories. This
supports their adequacy for the present dataset. It does not establish that the
taxonomy is complete for other languages, repositories, or development settings.
To assess whether the six categories can be applied reliably, two developers
independently categorized all 111 assumptions of the curated subset and then
reconciled their disagreements to consensus (rater background and protocol in
Appendix~\ref{app:benchmark}).
Agreement before reconciliation was strong (Cohen's $\kappa=0.85$, 88.3\% raw
agreement), indicating that the taxonomy can be applied consistently by a second
developer rather than being idiosyncratic to its author (details in
\S\ref{sec:threats}).

\begin{table}[t]
\caption{Taxonomy of implicit assumptions in LLM-generated code.
         The last column is the percentage of the 180 tasks that contain at
         least one instance of the category. Because a task usually spans
         several categories, this column does not sum to 100\% and is distinct
         from the per-assumption shares in \cref{tab:benchstats}.}
\label{tab:taxonomy}
\resizebox{\columnwidth}{!}{%
\begin{tabular}{@{}llp{3.5cm}r@{}}
\toprule
\textbf{ID} & \textbf{Category} & \textbf{Description} & \textbf{\% of tasks} \\
\midrule
T1 & Input Validation & Input format, encoding, boundary checks, type coercion & 86\% \\
T2 & Data Format      & Return type, output structure, serialization, nullability & 75\% \\
T3 & Error Policy     & Exception types, return sentinels, error propagation     & 81\% \\
T4 & Persistence      & Storage backend, schema, consistency guarantees          & 4\% \\
T5 & Performance      & Algorithm complexity, caching, data structure choices    & 76\% \\
T6 & Security         & Auth, cryptography, rate limiting, access control        & 3\% \\
\bottomrule
\end{tabular}}
\end{table}

\paragraph{Category characterization.}
\textbf{T1 (Input Validation)} is the most frequent category (86\% of tasks)
and captures assumptions about what the caller
guarantees: that a list argument is non-empty, that a string is valid UTF-8,
or that an integer falls within a documented range: silent preconditions whose
violation causes runtime errors rather than wrong results.
\textbf{T2 (Data Format)} is nearly as common (75\% of tasks) because
most functions must decide how to represent their output (as an index
or a value, as a dict or a dataclass, as \code{None} or an empty list on
failure) without any prompt guidance.
\textbf{T3 (Error Policy)} covers the choice of \emph{how} failures are
communicated: raised exceptions vs.\ sentinel returns vs.\ silent defaults.
A function returning \code{False} on bad input versus raising \code{ValueError}
are semantically equivalent in isolation but diverge sharply under a caller that
inspects the return type.
\textbf{T4 (Persistence)} applies when a function writes or reads state: the
choice between an in-memory dict and a database, or between append and overwrite,
is invisible in the function signature but consequential for any multi-session caller.
\textbf{T5 (Performance)} reflects algorithm or data-structure trade-offs (hash
map vs.\ sort-and-scan, eager vs.\ lazy evaluation) that determine asymptotic
complexity without appearing in the function signature.
\textbf{T6 (Security)} is the least frequent (3\%) but highest-stakes category:
a password stored as an MD5 hash passes all functional tests yet violates every
modern security policy.

\paragraph{Distribution pattern.}
Input validation, error policy, performance, and data format each appear in
roughly three-quarters or more of tasks, because most non-trivial functions
must decide how to treat their inputs, signal failure, structure output, and
trade off performance.
T4 (Persistence, 4\%) and T6 (Security, 3\%) are rare because persistence- and
security-sensitive tasks are underrepresented in general coding prompts. Their low frequency should \emph{not} be interpreted
as low importance: a single T6 assumption (e.g., weak cryptography) can
invalidate an entire system.
This skew motivates our per-category breakdown in \cref{tab:rq1breakdown}:
aggregate $F_1$ alone would mask systematic failures on low-frequency but
critical categories.

\section{The \sysname Framework}
\label{sec:framework}

\subsection{Architecture Overview}
\label{sec:architecture}

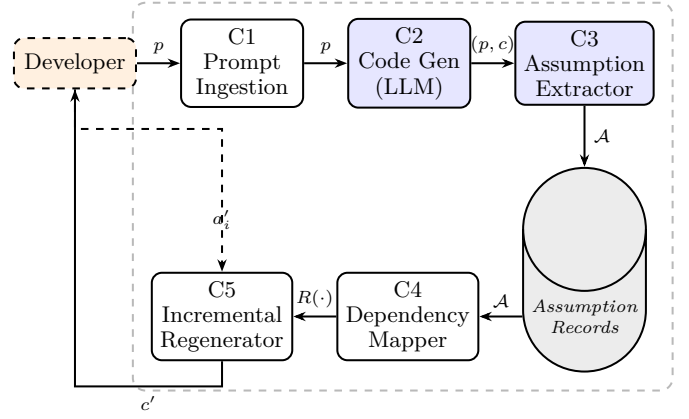
\begin{figure}[t]
  \centering
  \resizebox{\columnwidth}{!}{%
  \begin{tikzpicture}[
    node distance=0.45cm and 0.6cm,
    comp/.style={draw, rounded corners=4pt, minimum width=1.7cm, minimum height=0.7cm,
                 align=center, font=\small, fill=white, thick},
    llm/.style={comp, fill=blue!10},
    store/.style={draw, cylinder, shape border rotate=90,
                  minimum width=1.4cm, minimum height=0.55cm,
                  align=center, font=\scriptsize\itshape, fill=gray!15, thick},
    dev/.style={comp, fill=orange!12, dashed, thick},
    arr/.style={-{Stealth[scale=0.75]}, thick},
    lbl/.style={font=\scriptsize\itshape}
  ]
    \node[dev]     (dev)  {Developer};
    \node[comp, right=of dev]  (c1) {C1\\Prompt\\Ingestion};
    \node[llm,  right=of c1]   (c2) {C2\\Code Gen\\(LLM)};
    \node[llm,  right=of c2]   (c3) {C3\\Assumption\\Extractor};

    \node[store, below=0.85cm of c3]  (store) {Assumption\\Records};
    \node[comp,  left=0.6cm of store] (c4)    {C4\\Dependency\\Mapper};
    \node[comp,  left=0.6cm of c4]    (c5)    {C5\\Incremental\\Regenerator};

    \draw[arr] (dev) -- node[lbl,above]{$p$} (c1);
    \draw[arr] (c1)  -- node[lbl,above]{$p$} (c2);
    \draw[arr] (c2)  -- node[lbl,above]{$(p,c)$} (c3);

    \draw[arr] (c3)    -- node[lbl,right]{$\mathcal{A}$} (store);
    \draw[arr] (store) -- node[lbl,above]{$\mathcal{A}$} (c4);
    \draw[arr] (c4)    -- node[lbl,above]{$R(\cdot)$}    (c5);

    \draw[arr] (c5.south) -- ++(0,-0.35)
               -| node[lbl, near start, below]{\scriptsize $c'$} (dev.south);

    \draw[arr,dashed] (dev.south) -- ++(0,-0.55)
               -| node[lbl, near end, below]{\scriptsize $a_i'$} (c5.north);

    \begin{scope}[on background layer]
      \node[draw=gray!55, dashed, rounded corners=5pt, thick,
            inner sep=0.25cm,
            fit=(c5)(c4)(store)(c3)] {};
    \end{scope}
  \end{tikzpicture}
  }%
  \caption{\sysname system architecture.  Boxes denote components. Arrows denote
           data flow.  The dashed region marks the developer interaction loop.}
  \label{fig:architecture}
\end{figure}

\sysname consists of five loosely-coupled components connected by a structured data model
(\cref{fig:architecture}).

\begin{description}[leftmargin=1em, noitemsep]
  \item[\textbf{C1: Prompt Ingestion.}]  Receives the developer's natural-language
       prompt $p$, optionally enriched with project-level context (language, framework,
       coding style guide).
  \item[\textbf{C2: Code Generator.}]  Issues $p$ to the backbone LLM and produces
       the initial code artifact $c$.  Any instruction-tuned LLM (GPT-4o,
       Qwen2.5-Coder-7B, or similar) may serve as the backbone.
  \item[\textbf{C3: Assumption Extractor.}]  Given $(p, c)$, applies a structured
       extraction prompt (\S\ref{sec:extractor}) to enumerate $\mathcal{A}(p,c)$ as a
       list of \emph{AssumptionRecord} objects (\S\ref{sec:schema}).
  \item[\textbf{C4: Dependency Mapper.}]  Links each assumption to the code regions
       it governs via a lightweight AST-level analysis (\S\ref{sec:dependency}).
  \item[\textbf{C5: Incremental Regenerator.}]  When the developer revises an
       assumption $a_i$, regenerates only the code regions $R(a_i)$ while preserving
       all other regions (\S\ref{sec:regeneration}).
\end{description}

\Cref{alg:assumptionminer} summarizes the end-to-end procedure.
Components C3--C5 correspond to its three phases. The remainder of this
section details each in turn.

\begin{algorithm}[t]
\caption{\sysname: assumption extraction, dependency mapping, and incremental
         regeneration.}
\label{alg:assumptionminer}
\begin{algorithmic}[1]
\REQUIRE prompt $p$, generated code $c$; category-to-node-type map $\mathcal{M}$;
         narrowing threshold $\theta$; context width $k$
\ENSURE assumption records $A$ with linked code regions; revised code on edit
\STATE \COMMENT{Phase 1: Assumption extraction (\S\ref{sec:extractor})}
\STATE $r \leftarrow \textsc{Llm}(\textsc{ExtractPrompt}(p, c))$
       \COMMENT{single two-phase prompt, temperature $0$}
\STATE $A \leftarrow \textsc{ParseRecords}(r)$
       \COMMENT{each $a$: description, category $\in$ T1--T6, severity, alternatives}
\STATE \COMMENT{Phase 2: Dependency mapping (\S\ref{sec:dependency})}
\STATE $(\mathcal{T}, N) \leftarrow \textsc{TreeSitterParse}(c)$
       \COMMENT{$N[t]$: AST nodes of type $t$}
\FORALL{$a \in A$}
    \STATE $K \leftarrow \textsc{Keywords}(a)$;\quad $R \leftarrow \emptyset$
    \FORALL{node type $t \in \mathcal{M}[a.\mathrm{category}]$ and node $n \in N[t]$}
        \IF{some $\kappa \in K$ occurs in $\textsc{Text}(n)$}
            \IF{$\textsc{Span}(n) \leq \theta$}
                \STATE $(s,e) \leftarrow (n.\mathrm{start},\, n.\mathrm{end})$
            \ELSE
                \STATE $(s,e) \leftarrow \textsc{NarrowToKeywordLines}(n, K)$
                       \COMMENT{fall back to full span if no keyword line}
            \ENDIF
            \STATE $R \leftarrow R \cup \{\textsc{CodeRef}(s, e, t)\}$
        \ENDIF
    \ENDFOR
    \STATE $a.\mathrm{code\_refs} \leftarrow R$
\ENDFOR
\STATE \COMMENT{Phase 3: Incremental regeneration when the user revises $a\in A$
       to a new description $d'$ (\S\ref{sec:regeneration})}
\FORALL{$\rho \in \textsc{SortByStartDesc}(a.\mathrm{code\_refs})$}
    \STATE $x \leftarrow \textsc{ContextWindow}(c, \rho, k)$
           \COMMENT{$\pm k$ lines around region $\rho$}
    \STATE $y \leftarrow \textsc{Llm}(\textsc{RegenPrompt}(a.\mathrm{description}, d', x))$
    \STATE $c \leftarrow \textsc{Splice}(c, \rho, \textsc{ExtractCode}(y), k)$
           \COMMENT{reverse order keeps line offsets valid}
\ENDFOR
\RETURN $A$ and, on revision, the updated code $c$
\end{algorithmic}
\end{algorithm}

\subsection{Assumption Extraction}
\label{sec:extractor}

The core of \sysname is a two-phase extraction prompt (\cref{lst:prompt}).
Phase~1 asks the LLM to reason about what choices were \emph{made} in $c$ that go
beyond what $p$ requires, a chain-of-thought~\cite{wei2022cot} reasoning step.
Phase~2 asks the model to format its discoveries into the structured
\emph{AssumptionRecord} schema, enabling downstream programmatic processing.
The full prompt text is given in \cref{app:prompt}.

\begin{figure}[t]
\begin{lstlisting}[caption={Assumption extraction prompt template (abbreviated).
  \texttt{\{PROMPT\}} and \texttt{\{CODE\}} are replaced at runtime.},
  label={lst:prompt}, language={}]
SYSTEM: You are an expert software engineer auditing LLM-generated
code for implicit design decisions.

USER:
### Original Prompt
{PROMPT}

### Generated Code
{CODE}

### Task
1. REASON: List every design decision in the code that is NOT
   required by the prompt. For each, name a realistic alternative.
2. FORMAT: Return a JSON array of AssumptionRecord objects.
   Schema: {schema}
   Return ONLY valid JSON after the tag [RECORDS].

[RECORDS]
\end{lstlisting}
\end{figure}

\subsection{AssumptionRecord Schema}
\label{sec:schema}

Each extracted assumption is stored as an \emph{AssumptionRecord}
(\cref{lst:schema}).  The schema is designed for both human readability and
programmatic processing by C4 (dependency mapping) and C5 (incremental regeneration).

\begin{figure}[t]
\begin{lstlisting}[caption={AssumptionRecord JSON schema.},
  label={lst:schema}, language={}]
{
  "id":          "string",       // e.g. "A1"
  "category":    "T1|T2|T3|T4|T5|T6",
  "description": "string",      // natural-language statement
  "rationale":   "string",      // why LLM made this choice
  "alternatives": ["string"],   // >= 1 alternatives
  "code_refs": [{               // dependency info (C4 populates)
    "file":       "string",
    "start_line": "int",
    "end_line":   "int",
    "ast_node":   "string"      // optional AST node type
  }],
  "confidence": "float",        // [0, 1] extraction confidence
  "severity":   "low|medium|high"  // LLM-assigned; human may override
}
\end{lstlisting}
\end{figure}

\subsection{Assumption-Code Dependency Mapping}
\label{sec:dependency}

To support targeted incremental regeneration, \sysname must link each assumption to the
code regions it governs.  This is non-trivial: a single assumption (e.g., \emph{password
hashing algorithm}) may manifest across multiple non-contiguous lines and functions.

\begin{figure}[t]
  \centering
  \resizebox{0.97\columnwidth}{!}{%
  \begin{tikzpicture}[
    anode/.style={draw, circle, minimum size=0.65cm, font=\small\bfseries, thick},
    cnode/.style={draw, rounded corners=3pt, minimum width=2.4cm, minimum height=0.5cm,
                  align=left, font=\scriptsize, fill=gray!10, thick},
    arr/.style={-{Stealth[scale=0.7]}, gray!70, thick},
    lbl/.style={font=\scriptsize\itshape, gray!80}
  ]
    \node[anode, fill=red!15]    (a1) at (0, 2.2)  {$A_1$};
    \node[anode, fill=blue!15]   (a2) at (0, 1.1)  {$A_2$};
    \node[anode, fill=orange!20] (a3) at (0, 0.0)  {$A_3$};
    \node[anode, fill=green!15]  (a4) at (0,-1.1)  {$A_4$};
    \node[anode, fill=purple!12] (a5) at (0,-2.2)  {$A_5$};

    \node[font=\scriptsize, gray, anchor=east] at (-0.45, 2.2)  {T6};
    \node[font=\scriptsize, gray, anchor=east] at (-0.45, 1.1)  {T2};
    \node[font=\scriptsize, gray, anchor=east] at (-0.45, 0.0)  {T3};
    \node[font=\scriptsize, gray, anchor=east] at (-0.45,-1.1)  {T4};
    \node[font=\scriptsize, gray, anchor=east] at (-0.45,-2.2)  {T5};

    \node[cnode] (n1) at (3.5, 2.2)  {\texttt{def hash\_password()}};
    \node[cnode] (n2) at (3.5, 1.1)  {\texttt{token = uuid4()}};
    \node[cnode] (n3) at (3.5, 0.0)  {\texttt{return False}};
    \node[cnode] (n4) at (3.5,-1.1)  {\texttt{sessions = \{\}}};
    \node[cnode] (n5) at (3.5,-2.2)  {\texttt{def authenticate()}};

    \draw[arr] (a1) -- node[lbl, above=1pt] {0.92} (n1);
    \draw[arr] (a2) -- node[lbl, above=1pt] {0.88} (n2);
    \draw[arr] (a2) to[bend right=18] node[lbl, pos=0.35, right=2pt] {0.71} (n5);
    \draw[arr] (a3) -- node[lbl, above=1pt] {0.85} (n3);
    \draw[arr] (a4) -- node[lbl, above=1pt] {0.79} (n4);
    \draw[arr] (a5) -- node[lbl, below=1pt] {0.83} (n5);

    \node[font=\small\bfseries] at (0,  3.1) {Assumptions};
    \node[font=\small\bfseries] at (3.5, 3.1) {AST Nodes};
  \end{tikzpicture}
  }%
  \caption{Assumption-code dependency graph for the running example.
           Left: assumption nodes \Aone--\Afive (colored by category).
           Right: AST nodes in the generated code.
           Edge weight = dependency confidence.
           The graph is many-to-many: \Atwo governs two nodes, and
           \texttt{authenticate()} is governed by both \Atwo and \Afive.}
  \label{fig:dependency}
\end{figure}
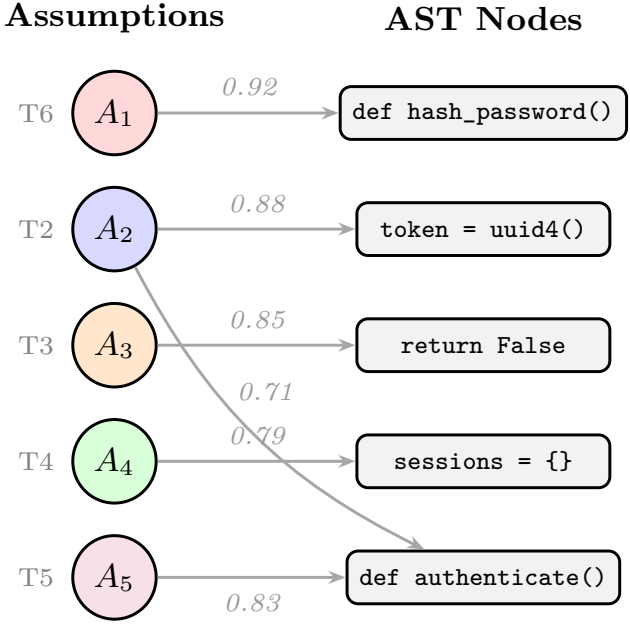

We construct an \emph{Assumption-Code Dependency Graph} $G = (V_a \cup V_c, E)$
(\cref{fig:dependency}), where $V_a$ is the set of assumptions, $V_c$ is the set of
AST nodes in $c$, and $(a_i, n_j) \in E$ if assumption $a_i$ governs node $n_j$.
Edge weights reflect confidence.

Edge construction proceeds in two passes:
\begin{enumerate}[noitemsep]
  \item \textbf{Keyword-overlap pass.}  We extract content words from the
        assumption's \emph{description} and \emph{alternatives} fields and
        tokenize the code identifiers.  AST nodes of the category-appropriate
        types (e.g., T3: \texttt{except\_clause}, \texttt{with\_statement},
        \texttt{call}; T5: \texttt{assignment}, \texttt{for\_statement};
        T6: \texttt{import\_statement}, \texttt{call}) are
        matched against node identifier text. A candidate is retained when
        any keyword from the assumption's \emph{description} or \emph{alternatives}
        appears as a substring in the node's identifier tokens.
        Optionally, an LLM call can replace keyword scoring with semantic
        region descriptions (\texttt{use\_llm=True}). We evaluate this variant in \S\ref{sec:rq2ablation}: a strong open-source mapper improves accuracy over the heuristic at additional inference cost.
  \item \textbf{AST refinement pass.}  We parse $c$ with the \code{tree-sitter}
        library and perform a bottom-up traversal from each candidate leaf node,
        stopping at the deepest ancestor that encloses \emph{all} candidate
        nodes for that assumption. This yields the minimal enclosing sub-tree.
\end{enumerate}

\subsection{Incremental Regeneration}
\label{sec:regeneration}

When a developer revises assumption $a_i$ to a new value $a_i'$, \sysname
identifies $R(a_i) = \{n \in V_c : (a_i, n) \in E\}$ and constructs a targeted
regeneration prompt containing only the affected sub-tree, the revised assumption,
and the surrounding context (at most $k$ lines above/below each region,
$k = 5$ in our experiments).
The remainder of $c$ is left unchanged.  This approach is analogous to
\emph{in-place surgical editing}~\cite{xia2023chatrepair}, but guided by explicit
assumption metadata rather than test-failure signals.

\section{Interactive Interface}
\label{sec:interface}

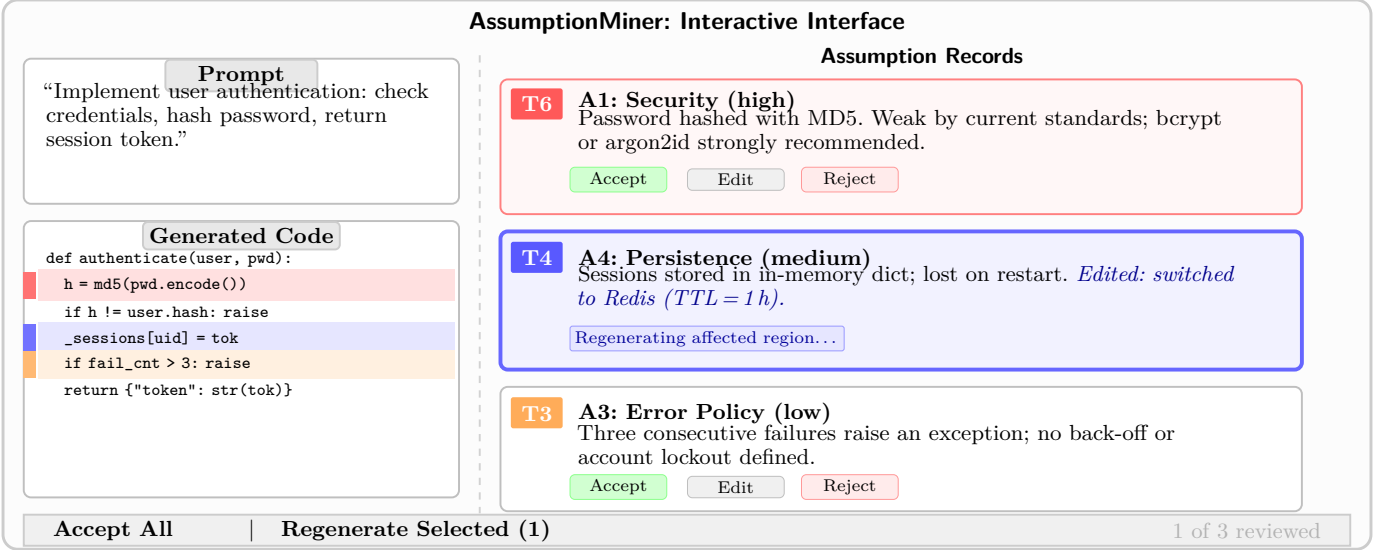
\begin{figure*}[t]
  \centering
  \resizebox{\textwidth}{!}{%
  \begin{tikzpicture}[
    sbox/.style={draw=gray!50, rounded corners=3pt, thick, fill=white},
    hdr/.style={draw=gray!40, rounded corners=2pt, fill=gray!18, thick,
                font=\small\bfseries, inner sep=2.5pt, anchor=north,
                minimum width=2.2cm},
    btn/.style={draw=gray!50, rounded corners=2pt, fill=gray!12,
                font=\scriptsize, inner sep=2pt, minimum width=1.4cm},
    gbtn/.style={btn, fill=green!18, draw=green!45},
    rbtn/.style={btn, fill=red!8, draw=red!40},
  ]
  \draw[draw=gray!40, rounded corners=6pt, very thick, fill=gray!2]
    (0, 1.65) rectangle (19.8, -6.30);
  \node[font=\normalsize\bfseries\sffamily, anchor=north] at (9.9, 1.60)
    {\textsc{AssumptionMiner}: Interactive Interface};
  \draw[sbox] (0.3, 0.80) rectangle (6.6, -1.30);
  \node[hdr]  at (3.45, 0.80) {Prompt};
  \node[font=\small, text width=5.8cm, anchor=north west] at (0.50, 0.58)
    {``Implement user authentication: check credentials, hash password, return session token.''};
  \draw[sbox] (0.3, -1.55) rectangle (6.6, -5.55);
  \node[hdr]  at (3.45, -1.55) {Generated Code};
  \node[anchor=west, font=\ttfamily\scriptsize, text width=5.8cm] at (0.50,-2.10)
    {\texttt{def authenticate(user, pwd):}};
  \node[anchor=west, font=\ttfamily\scriptsize, fill=red!12, text width=5.8cm] at (0.50,-2.48)
    {\texttt{\phantom{xx}h = md5(pwd.encode())}};
  \node[anchor=west, font=\ttfamily\scriptsize, text width=5.8cm] at (0.50,-2.86)
    {\texttt{\phantom{xx}if h != user.hash: raise}};
  \node[anchor=west, font=\ttfamily\scriptsize, fill=blue!10, text width=5.8cm] at (0.50,-3.24)
    {\texttt{\phantom{xx}\_sessions[uid] = tok}};
  \node[anchor=west, font=\ttfamily\scriptsize, fill=orange!12, text width=5.8cm] at (0.50,-3.62)
    {\texttt{\phantom{xx}if fail\_cnt > 3: raise}};
  \node[anchor=west, font=\ttfamily\scriptsize, text width=5.8cm] at (0.50,-4.00)
    {\texttt{\phantom{xx}return \{"token": str(tok)\}}};
  \draw[fill=red!55, draw=none] (0.30,-2.67) rectangle (0.48,-2.29);
  \draw[fill=blue!55, draw=none] (0.30,-3.43) rectangle (0.48,-3.05);
  \draw[fill=orange!55, draw=none] (0.30,-3.81) rectangle (0.48,-3.43);
  \draw[draw=gray!35, dashed, thick] (6.9, 0.85) -- (6.9, -5.78);
  \node[font=\small\bfseries\sffamily, anchor=south] at (13.3, 0.55)
    {Assumption Records};
  \draw[draw=red!50, rounded corners=3pt, thick, fill=red!3]
    (7.2, 0.50) rectangle (18.8, -1.45);
  \draw[fill=red!65, draw=none, rounded corners=1pt]
    (7.35,-0.08) rectangle (8.10, 0.37);
  \node[font=\small\bfseries, text=white] at (7.73, 0.15) {T6};
  \node[font=\small\bfseries, anchor=west] at (8.20, 0.18) {A1: Security (high)};
  \node[font=\small, anchor=west, text width=9.5cm] at (8.20,-0.25)
    {Password hashed with MD5.  Weak by current standards;
     bcrypt or argon2id strongly recommended.};
  \node[gbtn, anchor=west] at (8.20,-0.95) {Accept};
  \node[btn,  anchor=west] at (9.90,-0.95) {Edit};
  \node[rbtn, anchor=west] at (11.55,-0.95) {Reject};
  \draw[draw=blue!60, rounded corners=3pt, ultra thick, fill=blue!5]
    (7.2, -1.70) rectangle (18.8, -3.70);
  \draw[fill=blue!65, draw=none, rounded corners=1pt]
    (7.35,-2.30) rectangle (8.10, -1.85);
  \node[font=\small\bfseries, text=white] at (7.73,-2.08) {T4};
  \node[font=\small\bfseries, anchor=west] at (8.20,-2.10) {A4: Persistence (medium)};
  \node[font=\small, anchor=west, text width=9.5cm] at (8.20,-2.52)
    {Sessions stored in in-memory dict; lost on restart.
     \textcolor{blue!55!black}{\textit{Edited: switched to Redis (TTL\,=\,1\,h).}}};
  \node[font=\scriptsize, draw=blue!40, fill=blue!10, rounded corners=1pt,
        inner sep=2pt, text=blue!60!black, anchor=west] at (8.20,-3.25)
    {Regenerating affected region\ldots};
  \draw[draw=gray!50, rounded corners=3pt, thick, fill=white]
    (7.2, -3.95) rectangle (18.8, -5.75);
  \draw[fill=orange!65, draw=none, rounded corners=1pt]
    (7.35,-4.55) rectangle (8.10, -4.10);
  \node[font=\small\bfseries, text=white] at (7.73,-4.33) {T3};
  \node[font=\small\bfseries, anchor=west] at (8.20,-4.33) {A3: Error Policy (low)};
  \node[font=\small, anchor=west, text width=9.5cm] at (8.20,-4.78)
    {Three consecutive failures raise an exception; no back-off or account lockout defined.};
  \node[gbtn, anchor=west] at (8.20,-5.40) {Accept};
  \node[btn,  anchor=west] at (9.90,-5.40) {Edit};
  \node[rbtn, anchor=west] at (11.55,-5.40) {Reject};
  \draw[draw=gray!40, fill=gray!12, thick]
    (0.3,-5.80) rectangle (19.5,-6.25);
  \node[font=\small\bfseries, anchor=west] at (0.60,-6.025) {Accept All};
  \node[font=\small] at (3.60,-6.025) {|};
  \node[font=\small\bfseries, anchor=west] at (3.90,-6.025) {Regenerate Selected (1)};
  \node[font=\small, text=gray!60, anchor=east] at (19.2,-6.025) {1 of 3 reviewed};
  \end{tikzpicture}}%
  \caption{\sysname interactive interface.  (1) Highlighted code regions correspond
           to extracted assumptions.  (2) The assumption panel lets developers inspect
           and edit each assumption record.  (3) The revision panel applies changes
           through incremental regeneration.}
  \label{fig:ui}
\end{figure*}

The \sysname interface (\cref{fig:ui}) is a web-based IDE panel implemented in
React that augments a standard code editor.  Developers interact through three
primary actions:

\begin{description}[noitemsep, leftmargin=1em]
  \item[\textbf{Inspect.}]  Each AssumptionRecord is a severity-colored card
       (red/amber/green). Hovering highlights its code region in the editor.
  \item[\textbf{Edit.}]  Developers modify the \emph{description} or pick an
       \emph{alternative}. Changes are staged, not yet applied.
  \item[\textbf{Regenerate.}]  Pressing \emph{Regenerate} triggers C5 for all
       staged edits, topologically ordered over the dependency graph.
\end{description}

\textbf{Implementation.}
\sysname is implemented as an open-source Python package
(\texttt{assumption\_miner}), one module per component (C1--C5), with the
interface served as a web application and a unit-test suite of 119 tests.
Appendix~\ref{app:software} details the module layout, experiment entry
points, and extensibility. The package and all experiment scripts are part of
the released artifact.

\section{The \sysname Benchmark}
\label{sec:benchmark}

To enable systematic, reproducible evaluation of assumption mining methods, we
construct the \sysname benchmark (\cref{tab:benchstats}).

\textbf{Task collection.}  The benchmark consists of 180 programming tasks under
ambiguous or underspecified prompts.
Thirty tasks are novel, spanning diverse domains (data structures, algorithms,
file I/O, networking, system utilities), stress-testing the underrepresented
T4--Persistence and T6--Security categories, and selected to be
out-of-distribution with respect to HumanEval~\cite{chen2021codex} and
MBPP~\cite{austin2021mbpp}, which supply the remaining 150 tasks.

\textbf{Annotation.}  All 180 tasks use GPT-4o~\cite{openai2024gpt4o} to extract the reference
assumption list with the same structured prompt as the extractor
(\S\ref{sec:extractor}), followed by programmatic validation (schema
conformance, category-vocabulary checks, and deduplication).
The 30 novel tasks form the \emph{curated subset}: their prompts were designed
specifically for this benchmark, and they additionally carry code-region
references (line ranges) used for RQ2.
Since the reference annotations are LLM-derived, two developers, each with at
least five years of professional software-development experience in industry,
independently verified the curated subset (111 assumptions), spending roughly
four hours each, and reconciled to consensus:
they agreed that 110 of 111 are valid implicit assumptions (99.1\% raw
agreement, $\kappa=0.66$) and agreed on category at $\kappa=0.85$
(\S\ref{sec:threats}, Appendix~\ref{app:benchmark}).
Adjudication then removed two invalid entries and corrected eight category
labels, yielding the 109 human-verified assumptions used throughout the paper.
Because GPT-4o produced the reference annotations, results for the GPT-4o
backbone carry a self-consistency risk, marked~$\dagger$ in
\cref{tab:rq1main}. The open-source backbones (Qwen32B, DeepSeek) share no
weights or provider with the annotation model, though their scores still
measure agreement with the reference.
All tasks share the six-category schema (\cref{tab:taxonomy}).

\textbf{Benchmark statistics} are summarized in \cref{tab:benchstats}.
Full benchmark construction details, including prompt selection criteria and
the annotation resolution process, are provided in \cref{app:benchmark}.

\begin{table}[t]
\caption{Benchmark composition. The 180 tasks split into the 30-task curated
         subset, which carries human-verified code-region references (used for
         RQ2), and the 150-task extended tier drawn from HumanEval/MBPP.
         All headline results (\cref{tab:rq1main,tab:rq1breakdown}, RQ3) use
         all 180 tasks.}
\label{tab:benchstats}
\resizebox{\columnwidth}{!}{%
\begin{tabular}{@{}lrrr@{}}
\toprule
\textbf{Metric} & \textbf{Curated subset} & \textbf{Extended tier} & \textbf{All (180)} \\
\midrule
Tasks                             & 30    & 150   & 180 \\
Total assumptions                 & 109   & 567   & 676 \\
Mean assumptions / task           & 3.6   & 3.8   & 3.8 \\
Mean lines of code / task         & 11.0  & 10.6  & 10.7 \\
\midrule
\multicolumn{4}{@{}l}{\textit{Category breakdown: count and \% of each tier's assumptions (columns sum to 100\%)}} \\
T1 Input Validation               & 20 (18\%) & 183 (32\%) & 203 (30\%) \\
T2 Data Format                    & 31 (28\%) & 112 (20\%) & 143 (21\%) \\
T3 Error Policy                   & 24 (22\%) & 142 (25\%) & 166 (25\%) \\
T4 Persistence                    & 8 (7\%)   & 0 (0\%)    & 8 (1\%)   \\
T5 Performance                    & 21 (19\%) & 130 (23\%) & 151 (22\%) \\
T6 Security                       & 5 (5\%)   & 0 (0\%)    & 5 (1\%)   \\
\bottomrule
\end{tabular}}
\end{table}

In the curated subset, 67\% of code-region references are single-line (mean
span: 2.0 lines), suggesting assumptions typically manifest at specific
statements. This motivates our IoU threshold of 0.5 for RQ2.
All T4 (Persistence) and T6 (Security) assumptions in the benchmark occur in
the curated subset: the 150 HumanEval/MBPP tasks of the extended tier contain
none, so the novel tasks provide the benchmark's only coverage of these two
rare but high-stakes categories.
(Note the two frequency views: \cref{tab:taxonomy} reports the share of
\emph{tasks} containing $\geq 1$ instance of a category over all 180 tasks,
whereas the breakdown above reports each category's share of the
\emph{assumptions} in each tier.)

\textbf{Metrics.}
We define three extraction metrics over $\mathcal{A}(p,c)$ and a reference set
$\mathcal{A}^*(p,c)$:

\begin{description}[noitemsep, leftmargin=1em]
  \item[\AR (Assumption Recall):] $|\mathcal{A} \cap \mathcal{A}^*| \;/\; |\mathcal{A}^*|$
  \item[\AP (Assumption Precision):] $|\mathcal{A} \cap \mathcal{A}^*| \;/\; |\mathcal{A}|$
  \item[\AFone (Assumption $F_1$):] Harmonic mean of \AR and \AP
\end{description}

\textbf{Matching.} We report extraction quality under two matching rules of
increasing strictness.
\emph{Category-level (lenient)} matching counts a predicted assumption as
correct when its category equals that of an unused reference assumption for the
same task, with multiplicity (if the extractor proposes two T3 assumptions and
the reference contains one, only one is a true positive).
It answers \emph{``did the extractor notice a decision of this kind?''} and is
the basis of \AR, \AP, and \AFone. Being category-only, it is deliberately
lenient and comparable to how prior work scores taxonomy coverage.
\emph{Decision-level (strict)} matching additionally requires the predicted
\emph{description} to refer to the same decision as the reference: a match
holds only when categories agree \emph{and} the description similarity
$\text{sim}(d(a),d(a^*)) \geq \tau$.
It answers the harder question \emph{``did the extractor identify the actual
decision?''}
We compute $\text{sim}$ as the cosine similarity of sentence embeddings (a
deterministic, reproducible measure) and calibrate $\tau$ against human
same-decision judgments on a sample of matched pairs
(\S\ref{sec:rq1}, \texttt{scripts/run\_rq1\_semantic.py}). We deliberately
avoid an LLM judge, which would reintroduce the very LLM-grades-LLM circularity
we control for elsewhere.
For RQ2 dependency links, we use line-level IoU:
$\text{IoU}(\hat{R}, R^*) = |\hat{R} \cap R^*| \,/\, |\hat{R} \cup R^*|$;
a link is correct when IoU $\geq 0.5$.
This threshold is motivated by the reference annotation spans, which are
predominantly 1--3 lines. IoU$\geq$0.5 requires the predicted region to overlap
with at least half the annotated span.

\section{Evaluation}
\label{sec:evaluation}

We evaluate \sysname through three research questions.  RQ1 assesses whether
the system recovers implicit decisions from generated code. RQ2 assesses
whether it localizes those decisions to the relevant implementation region, and RQ3 assesses whether those links support more focused revisions.  This
sequence separates the three claims required by the approach: identifying a
decision, tracing it to code, and using the trace during change.

\subsection{Experimental Setup}
\label{sec:setup}

\textbf{Backbone LLMs.}  We evaluate \sysname with four backbone LLMs:
GPT-4o~\cite{openai2024gpt4o} (OpenAI API, model alias \code{gpt-4o}),
Qwen2.5-Coder-32B-Instruct (hereafter Qwen32B),
DeepSeek-Coder-V2-Lite-Instruct (hereafter DeepSeek), and
Qwen2.5-Coder-7B-Instruct (hereafter Qwen-7B). The latter
three are served locally via vLLM~\cite{kwon2023vllm} on a single NVIDIA RTX
PRO 6000 Blackwell workstation GPU (96\,GB of VRAM).
All are evaluated on the full $n=180$ benchmark tasks.
We additionally report a Conf-Weighted (\textbf{CW}) ensemble combining
Qwen2.5-Coder-32B (high-recall backbone) and DeepSeek-Coder-V2-Lite
(high-precision backbone), averaging per-assumption confidence scores and
thresholding at 0.5.
Results for all backbones are reported in \cref{tab:rq1main}.

\textbf{Baselines.}
\begin{itemize}[noitemsep]
  \item \textit{Comment Extraction (CE):} Extracts design decisions from
        code comments and docstrings only, and requires no LLM call.
  \item \textit{Knowledge-Based Pattern Extraction (KBE):} Mines Python AST
        structure and code idioms (return sentinels, exception breadth, hash-map
        use) against a hand-built rule base to infer implicit design choices, and runs entirely offline without an LLM.
\end{itemize}

We also considered three additional chain-of-thought/prompting baselines during
development: \emph{Direct Generation (DG)}, which asks the LLM to list assumptions
without the structured two-phase prompt; \emph{Clarifying Questions (CQ)}, which
prompts the LLM to generate questions it would ask before coding; and
\emph{Chain-of-Thought (CoT)} prompting~\cite{wei2022cot}, which elicits reasoning
before extraction.
In preliminary experiments on the 30 curated-subset tasks, DG reached $F_1=0.41$
(vs.\ 0.59 for the two-phase prompt), CQ reached $F_1=0.37$ (it generates
hypothetical questions rather than identifying actual code choices), and CoT
reached $F_1=0.56$.
None exceeded the structured two-phase prompt in either precision or recall. We therefore use the structured prompt as the single \sysname extraction method
and report only CE and KBE as offline baselines in the main tables.

\textbf{Dependency baselines (RQ2):}
\begin{itemize}[noitemsep]
  \item \textit{Keyword Heuristic (KH):} Token-overlap between assumption description
        and code tokens, with no AST information.
  \item \textit{Full-File Dependency (FF):} Trivially assigns every assumption to the
        entire file (upper bound on recall, lower bound on precision).
\end{itemize}

\textbf{Regeneration baselines (RQ3):}
\begin{itemize}[noitemsep]
  \item \textit{Full Regeneration (FR):} Re-generates the entire function from scratch
        with the revised assumption in the prompt.
  \item \textit{In-Place LLM Edit (IPE):} Passes the full code + revision instruction to
        the LLM and asks it to apply the change~\cite{xia2023chatrepair}.
  \item \textit{Manual Reprompt (MR):} Developer manually rewrites the prompt. LLM generates fresh code.
\end{itemize}

\subsection{RQ1: Extraction Quality}
\label{sec:rq1}

RQ1 asks whether \sysname can recover the implicit assumptions that a
developer would otherwise need to infer through code review.  We evaluate
extraction against the benchmark reference annotations on all 180 tasks.  We
first compare methods using category-level precision, recall, and $F_1$. This
measures whether a method identifies a decision of the appropriate kind.
We then examine performance by category and report a stricter decision-level
match to distinguish category coverage from description fidelity.  Because
GPT-4o created the initial reference annotations, we treat the open-source
backbones (Qwen32B, DeepSeek, Qwen-7B, and their CW ensemble), which share no
weights or provider with the annotation model, as the primary cross-model
evidence, and report the GPT-4o result separately (marked~$\dagger$).

\cref{tab:rq1main} presents the aggregate comparison.  It establishes the
overall extraction result before \cref{tab:rq1breakdown} examines where the
methods succeed or fail across assumption categories.

\begin{table}[t]
\caption{RQ1 assumption-extraction quality on the full benchmark ($n=180$, 676
         assumptions). Best per column in \textbf{bold}.
         $\dagger$\,GPT-4o is both annotator and evaluator here (it produced all
         reference annotations), so its scores may be inflated by
         self-consistency. The open-source backbones are fully cross-model.
         CW\,=\,Conf-Weighted cross-model ensemble. The strict
         \AFone{} column is explained in \S\ref{sec:rq1}.}
\label{tab:rq1main}
\resizebox{\columnwidth}{!}{%
\begin{tabular}{@{}lrrrr@{}}
\toprule
\textbf{Method} & \AR & \AP & \AFone & \AFone (strict) \\
\midrule
\multicolumn{5}{@{}l}{\textit{Offline baselines (no LLM, $n=180$)}} \\
Comment Extraction (CE)  & 0.176 & 0.312 & 0.225 & 0.079 \\
Pattern Extraction (KBE) & 0.044 & 0.476 & 0.081 & 0.019 \\
\midrule
\multicolumn{5}{@{}l}{\textit{\sysname (ours, $n=180$)}} \\
\sysname (GPT-4o)$^\dagger$                      & \textbf{0.905} & 0.864 & \textbf{0.884} & \textbf{0.806} \\
\sysname (Qwen2.5-Coder-32B)                     & 0.753 & 0.797 & 0.774 & 0.654 \\
\sysname (DeepSeek-Coder-V2-Lite)                & 0.732 & \textbf{0.814}          & 0.771 & 0.565 \\
\sysname (Qwen2.5-Coder-7B)                      & 0.468 & 0.810          & 0.593 & 0.463 \\
\midrule
\multicolumn{5}{@{}l}{\textit{Ensemble (ours, $n=180$, cross-model)}} \\
\sysname Ens. (Qwen32B+DeepSeek, CW)             & 0.823 & 0.810          & 0.816 & 0.662 \\
\bottomrule
\end{tabular}}

{\footnotesize\vspace{2pt}
Bootstrap 95\% CI over per-task $F_1$ (10{,}000 resamples of the 180 tasks):
GPT-4o$^\dagger$ $[0.866, 0.910]$, Qwen32B $[0.745, 0.794]$,
DeepSeek $[0.737, 0.785]$, Qwen-7B $[0.553, 0.608]$.}
\end{table}

The aggregate values do not reveal whether performance is concentrated in
common, easy-to-observe decisions.  We therefore break $F_1$ down by taxonomy
category in \cref{tab:rq1breakdown}. The complete precision and recall values
are deferred to Appendix~\ref{app:tables}.

\begin{table*}[t]
\caption{RQ1 per-category \AFone ($n=180$). Best cross-model $F_1$ per category in \textbf{bold}
         (GPT-4o$^\dagger$ produced the reference annotations).
         CE $F_1=0.363$ on T2 (return-type docstrings) but near-zero on all other categories.}
\label{tab:rq1breakdown}
\resizebox{\textwidth}{!}{%
\begin{tabular}{@{}lrrrrrrr@{}}
\toprule
\textbf{Method} & \textbf{T1} & \textbf{T2} & \textbf{T3} & \textbf{T4} & \textbf{T5} & \textbf{T6} & \textbf{All} \\
\midrule
Comment Extraction (CE)           & 0.171 & 0.363 & 0.058 & 0.000 & 0.116 & 0.000 & 0.225 \\
\sysname (GPT-4o)$^\dagger$       & 0.905 & 0.882 & 0.869 & 0.615 & 0.902 & 0.500 & 0.884 \\
\sysname (Qwen2.5-Coder-32B)      & 0.709 & 0.795 & \textbf{0.771} & \textbf{0.667} & \textbf{0.851} & 0.500 & 0.774 \\
\sysname (DeepSeek-Coder-V2-Lite) & \textbf{0.799} & 0.796 & 0.725 & 0.588 & 0.781 & \textbf{0.533} & 0.771 \\
\bottomrule
\end{tabular}}
\end{table*}

\textbf{Findings ($n=180$).}
The offline baselines provide little coverage because the decisions of interest
are seldom written in comments or directly encoded by a fixed pattern.  CE
reaches $F_1=0.225$, largely through documented return formats (T2), whereas
KBE reaches $F_1=0.081$ because its recall is only 0.044.

The strongest open-source result is the CW ensemble ($F_1=0.816$).
Qwen32B and DeepSeek are nearly tied overall ($F_1=0.774$ and 0.771), but
make complementary errors: Qwen32B recovers more assumptions and DeepSeek is
slightly more precise.  Their overlapping bootstrap intervals
(\cref{tab:rq1main}) indicate that this small difference is not meaningful.  GPT-4o obtains the highest score,
but we treat it as secondary evidence because it also produced the reference
annotations.

The per-category $F_1$ results (\cref{tab:rq1breakdown}) identify an important
boundary of the benchmark.
Data-format and performance decisions are comparatively easy to recover,
whereas the rare persistence and security decisions are harder.  All T4 and
T6 instances occur in the curated tasks, showing why conventional
HumanEval/MBPP-style prompts are insufficient for assessing these categories.

\textbf{Lenient vs.\ strict matching.}
The scores above use lenient category-level matching.
Under the strict \emph{decision-level} metric, which additionally requires the
predicted description to refer to the same decision as the reference
(\S\ref{sec:setup}), scores fall substantially.
We measure description similarity by sentence-embedding cosine
(\texttt{all-MiniLM-L6-v2}) and calibrate the match threshold against human
judgment: an author labeled 30 candidate (predicted, reference) pairs sampled
across the similarity range as same-decision or not (three boundary cases
adjudicated jointly), and $\tau=0.56$ best separates the labels
(24/30 = 80\% agreement, any $\tau$ between $0.55$ and $0.58$ performs
identically).
At $\tau=0.56$, the strict decision-level scores are
GPT-4o$^\dagger$ $F_1=0.806$ ($P=0.788$, $R=0.825$),
Qwen2.5-Coder-32B $F_1=0.654$, DeepSeek-Coder-V2-Lite $F_1=0.565$, and
Qwen2.5-Coder-7B $F_1=0.463$ (the strict column of \cref{tab:rq1main}), each
well below its lenient counterpart, and consistent with the purely lexical lower
bound ($F_1\approx0.60$ for Qwen2.5-Coder-32B), which penalizes paraphrases
that the embedding metric credits.
The CW ensemble attains the highest strict \AFone{} of any cross-model
configuration ($F_1=0.662$), mirroring its lenient lead and confirming that
merging a high-recall and a high-precision backbone helps under the stricter
criterion as well.
The offline baselines, by contrast, collapse to near-zero
(CE $F_1=0.079$, KBE $F_1=0.019$): the few assumptions they surface almost
never name the same decision as the reference.
To score the baselines and the ensemble strictly (neither emits per-assumption
descriptions natively), we reconstruct one description per prediction from its
extracted span (\texttt{scripts/build\_strict\_inputs.py}), which reproduces
each method's lenient prediction count exactly (CE 381, KBE 63, and ensemble
689 predictions, the last giving lenient $F_1=0.816$).
The gap is not a weakness of \sysname's framework but a property of the
benchmark: identifying \emph{which} implicit decision was made, not merely
that a decision of some category exists, remains difficult for current models,
so the benchmark is far from saturated and retains headroom to measure future
progress.

\noindent\textbf{RQ1 answer.}
Open-source models recover implicit assumptions substantially better than the
offline comment- and pattern-based baselines. The cross-model CW ensemble
reaches $F_1=0.816$, compared with 0.225 for the strongest offline baseline.
Exact decision matching remains more difficult than category-level matching,
especially for the rare security and persistence categories.

\subsection{RQ2: Dependency Accuracy}
\label{sec:rq2}

RQ2 asks whether the dependency mapper identifies the code region that
implements a recovered assumption. This question matters because localized
regeneration is meaningful only when the selected region is sufficiently
specific. Our primary analysis uses the 109 human-verified code-region
references in the curated subset. We also report a secondary analysis on all
676 references. Its additional 567 references are auto-labeled, so those
results measure agreement with the generated reference rather than accuracy
against ground truth. We report both threshold accuracy (IoU$\geq0.5$) and
mean IoU to distinguish selecting the right area from selecting a tight span.

\begin{table}[H]
\caption{RQ2: Dependency accuracy (Acc.\,=\,IoU$\geq$0.5; IoU\,=\,mean;
         Exact\,=\,exact-match rate; Calls\,=\,LLM calls per assumption).
         $n_C$\,=\,109 curated-subset references (primary);
         $n_F$\,=\,676 full set (adds 567 auto-labeled references; secondary).
         AM-LLM\,=\,\sysname with \texttt{use\_llm=True} and the given mapper
         model (\S\ref{sec:rq2ablation}). Both mappers are evaluated on the
         curated and full sets. Best per column in \textbf{bold}.}
\label{tab:rq2}
\resizebox{\columnwidth}{!}{%
\begin{tabular}{@{}lrrrrrr@{}}
\toprule
& \multicolumn{3}{c}{\textbf{Curated ($n_C$)}} & \multicolumn{2}{c}{\textbf{Full ($n_F$)}} & \\
\cmidrule(lr){2-4}\cmidrule(lr){5-6}
\textbf{Method} & \textbf{Acc} & \textbf{IoU} & \textbf{Exact} & \textbf{Acc} & \textbf{IoU} & \textbf{Calls} \\
\midrule
Keyword Heuristic (KH)      & 0.606 & 0.545 & 0.431 & 0.284 & 0.288 & 0 \\
Full-File (FF)              & 0.064 & 0.196 & 0.018 & 0.368 & 0.397 & 0 \\
\sysname (heuristic)        & 0.569 & 0.581 & \textbf{0.505} & 0.376 & 0.376 & 0 \\
AM-LLM (GPT-4o)             & 0.560 & 0.596 & 0.431 & 0.475 & \textbf{0.480} & 1 \\
AM-LLM (Qwen2.5-Coder-32B)  & \textbf{0.642} & \textbf{0.653} & \textbf{0.505} & \textbf{0.479} & 0.464 & 1 \\
\bottomrule
\end{tabular}}
\end{table}

\textbf{Findings.}
The curated subset is the primary result because its references were
human-verified.  On this set, \sysname and keyword matching have similar
threshold accuracy (56.9\% and 60.6\%), but AST narrowing yields the tightest
regions (mean IoU 0.581, versus 0.545 for keyword matching).  The whole-file
baseline performs poorly because these references usually identify a small
part of the implementation.
\Cref{tab:ablation} ablates the mapper's category-to-node-type map
(\S\ref{sec:dependency}): \emph{Full} is the shipped map, and
\emph{Original} is the map before development-time tuning added the
T3 \texttt{call}, T3 \texttt{with\_statement}, and
T5 \texttt{import\_from\_statement} types (its ``expansions''), together
worth $+2.7$pp.  Removing the base T5 \texttt{assignment} type, present in
both maps, costs the same $2.7$pp on its own.  Assignments (e.g.,
\texttt{cache = \{\}}) are where data-structure choices surface, so they
are important for localizing performance decisions.

The full-set result should be interpreted differently.  It includes 567
auto-labeled references for very short functions, where marking the entire
function is often sufficient.  In that setting \sysname improves on keyword
matching but is close to the whole-file baseline.  We therefore use the
curated-subset result to support claims about localization precision and report
the full set as a secondary robustness check.

\begin{table}[H]
\caption{RQ2 ablation: dependency mapper accuracy when removing one node type
         at a time from the full configuration (30 curated-subset tasks, 109 assumptions).}
\label{tab:ablation}
\resizebox{\columnwidth}{!}{%
\begin{tabular}{@{}lrr@{}}
\toprule
\textbf{Configuration} & \textbf{Acc.} & \textbf{$\Delta$ Acc.} \\
\midrule
Full (all expansions)                       & 0.569 & n/a \\
\quad$-$T3 \texttt{call}                   & 0.560 & $-$0.9pp \\
\quad$-$T3 \texttt{with\_statement}        & 0.560 & $-$0.9pp \\
\quad$-$T5 \texttt{import\_from\_statement}& 0.569 & 0.0pp \\
\quad$-$T5 \texttt{assignment}             & 0.541 & $-$2.7pp \\
Original (no expansions)                    & 0.541 & $-$2.7pp \\
\bottomrule
\end{tabular}}
\end{table}

We also asked whether the dependency map's accuracy carries through to code
adaptability: whether tasks with more accurate maps also receive smaller edits
in RQ3.
On the human-verified references this correlation is weak: splitting the 30
tasks by \sysname link accuracy (IoU$\geq$0.5) into high ($n=23$) and low
($n=7$) groups, mean edit distance rises only from 0.258 to 0.281
($1.09\times$), and In-Place LLM Edit shows a similar spread (0.264 vs.\
0.323).
We therefore do \emph{not} claim a strong mapping-accuracy-to-edit-size effect. The value of the dependency map is in enabling \emph{targeted} regeneration at
all (RQ3), not in a monotonic accuracy-edit relationship.
\S\ref{sec:rq2ablation} shows an LLM-guided mapper with a strong open-source
mapper can improve over this heuristic on the curated subset (the AM-LLM rows
of \cref{tab:rq2}), at the cost of one additional inference call per
assumption.

\noindent\textbf{RQ2 answer.}
On human-verified references, AST narrowing produces tighter localizations than
keyword matching, although their threshold accuracies are similar. An optional
LLM-guided mapper improves accuracy to 64.2\% at the cost of an additional
inference call. Whole-file localization is competitive only on the secondary
set of very short, auto-labeled functions.

\subsection{Mapper Ablation: Use-LLM vs.\ Heuristic}
\label{sec:rq2ablation}

We ablate whether replacing keyword-overlap scoring with an LLM-guided
localization step (\texttt{use\_llm=True}, \S\ref{sec:dependency}) improves
dependency accuracy.  We evaluate five mapper configurations on the 109
human-verified assumptions of the curated subset, scored against the consensus
reference regions: \sysname (heuristic), AM-LLM (Qwen32B), AM-LLM (GPT-4o),
KH, and FF.  All five appear in the $n_C$ columns of \cref{tab:rq2}, together
with each variant's per-assumption inference cost.

\textbf{Findings.}
The LLM-guided mapper with a Qwen2.5-Coder-32B backbone achieves \textbf{64.2\%}
accuracy (IoU $0.653$), the best of all mapper variants in \cref{tab:rq2},
exceeding the \sysname heuristic (56.9\%, IoU $0.581$) by 7.3~pp.
With a GPT-4o mapper, the LLM-guided variant scores 56.0\% (IoU $0.596$), on
par with the heuristic (within 1~pp on accuracy).
The result is therefore mapper-dependent: a strong open-source \emph{code}
model improves localization over keyword-narrowing plus AST filtering, while
GPT-4o does not, and either LLM variant incurs one additional inference call
per assumption.
Both LLM mappers lead on the secondary full set (\cref{tab:rq2}, $n_F$
columns): Qwen32B reaches 47.9\% accuracy (IoU $0.464$) and GPT-4o 47.5\%
(IoU $0.480$), versus 37.6\% for the heuristic and 36.8\% for whole-file
localization. Whereas the heuristic barely separates from whole-file
localization on these short auto-labeled functions, both LLM mappers keep a
clear margin, so on the full set the mapper choice matters less than
\emph{whether} an LLM mapper is used at all.
We keep the AST heuristic as \sysname's default for its zero inference cost
and competitive accuracy, and expose the LLM-guided mapper via the
\texttt{use\_llm} flag for accuracy-critical settings.
(An earlier version of this comparison, which scored noisy \emph{predicted}
extractor descriptions against the pre-correction references, favored the
heuristic, 45.5\% vs.\ 30.9--31.8\%. That reference set has since been
superseded by the human-verified consensus, so we report the numbers above
as primary.)

\subsection{RQ3: Code Adaptability}
\label{sec:rq3}

RQ3 asks whether assumption-guided regeneration reduces the scope of a code
change when a developer revises one decision. We use the full 180-task
benchmark with GPT-4o and repeat the \sysname configuration on the 30 novel
tasks with Qwen32B. Assumptions are extracted with C3
(\S\ref{sec:extractor}). No manual dependency annotation is required because
syntactic validity and edit distance are measured on the regenerated output.

\textbf{Assumption modification procedure.}
For each task we select the highest-severity AssumptionRecord and rewrite its
\emph{description} to the first entry in its \texttt{alternatives} field, producing a
(code, original-assumption, revised-assumption) triple.
The revised record is passed to the regenerator (\S\ref{sec:regeneration}) together with the
affected code sub-tree and $k=5$ context lines. Only the regions governed by the selected
assumption are updated.
This operationalization ensures the revision is semantically meaningful (a real alternative
that the LLM could have chosen) and reproducible (deterministic given the AssumptionRecord).

\emph{Worked example.}
For the authentication task of the running example, the selected record is
\Aone{} (T6): \emph{``Passwords are hashed with MD5, a cryptographically broken
algorithm''}, whose first alternative is \emph{``bcrypt.''}
The regenerator receives the revised record, the dependency-mapped sub-tree
(the \code{hash\_password} function body), and five context lines. It rewrites
the hashing call and its salt handling, leaving session, persistence, and
rate-limiting code untouched.
On this task the regenerated code differs from the original by a normalized
edit distance of 0.40, versus 0.85 for Full Regeneration and 0.82 for Manual
Reprompt, which rewrite the entire function.

\textbf{Metrics.}
\begin{itemize}[noitemsep]
  \item \emph{Syntactic Validity Rate (SVR)}: fraction of tasks where the
        revised code compiles (\code{py\_compile}).
        This is a necessary but not sufficient condition for functional
        correctness. It does \emph{not} verify that the revision implements the
        intended behavior, and full unit-test evaluation is future work
        (\S\ref{sec:threats}).
  \item \emph{Edit Distance}: character-level Levenshtein distance between
        original and revised code, normalized by the length of the longer
        string (lower = more targeted).
  \item \emph{Latency}: wall-clock time for the regeneration step in seconds.
\end{itemize}

\begin{table}[t]
\caption{RQ3: Code adaptability when one assumption is revised ($n=180$, GPT-4o, full benchmark).
         SVR = syntactic validity rate (\code{py\_compile});
         Edit Dist. = normalized Levenshtein (lower = more targeted).
         \sysname achieves lowest edit distance in all configurations.}
\label{tab:rq3}
\resizebox{\columnwidth}{!}{%
\begin{tabular}{@{}lrrr@{}}
\toprule
\textbf{Method} & \textbf{SVR} & \textbf{Edit Dist.} & \textbf{Lat. (s)} \\
\midrule
Manual Reprompt       & 1.000 & 0.574 & 1.79 \\
Full Regeneration     & 1.000 & 0.549 & 1.61 \\
In-Place LLM Edit     & 1.000 & 0.306 & 1.46 \\
\sysname (Incremental)& 0.833 & \textbf{0.244} & 2.40 \\
\midrule
\multicolumn{4}{@{}l}{\textit{Cross-model: \sysname (Qwen2.5-Coder-32B, $n=30$ novel tasks)}} \\
\sysname (Qwen2.5-Coder-32B, local)& 0.833 & \textbf{0.242} & 3.58 \\
\bottomrule
\end{tabular}}
\end{table}

\textbf{Findings.}
\sysname makes the smallest edits in both configurations (0.244 on the
full benchmark and 0.242 with Qwen32B on the novel tasks).  Its advantage is
substantial over full regeneration and manual reprompting, and smaller but
still present relative to in-place editing.  The bootstrap intervals for the
full benchmark do not overlap with those of any baseline.

This benefit has a clear trade-off: the targeted pipeline achieves 83.3\%
syntactic validity, whereas all non-targeted baselines compile in every case.
The error analysis attributes most failures to cascading edits, in which a
revision invalidates another linked region.  Thus, the result supports targeted
regeneration as a way to limit change scope, not as a substitute for validation
of the revised program.

\noindent\textbf{RQ3 answer.}
Assumption-guided regeneration reduces the size of a revision, with the lowest
edit distance in both configurations. It currently trades that precision for
lower syntactic validity (83.3\% versus 100\% for the non-targeted baselines),
so conflict detection and functional testing are necessary next steps.

\subsection{Error Analysis}
\label{sec:erroranalysis}

\cref{fig:erroranalysis} illustrates six representative failure modes: three
in extraction (1--3) and three in localization and regeneration (4--6).
For the extraction modes, \cref{tab:errorprofile} summarizes each backbone's
error profile, derived from the saved RQ1 predictions (per-category
false-negative and false-positive counts follow from per-category precision
and recall over the 676 reference assumptions).  Modes (4)--(6) draw on the
RQ2 and RQ3 runs (\cref{tab:rq2,tab:rq3}). The cases in
\cref{fig:erroranalysis} are illustrative.

\begin{table}[H]
\caption{Extraction error profile per backbone, derived from the saved RQ1
         predictions ($n=180$, 676 reference assumptions).}
\label{tab:errorprofile}
\centering
\resizebox{\columnwidth}{!}{%
\begin{tabular}{@{}lrrcc@{}}
\toprule
\textbf{Backbone} & \textbf{FN} & \textbf{FP} & \textbf{Top FN cat.} & \textbf{Top FP cat.} \\
\midrule
GPT-4o$^\dagger$        &  64 &  96 & T2 (36\%) & T3 (34\%) \\
Qwen2.5-Coder-32B       & 167 & 130 & T1 (49\%) & T3 (54\%) \\
DeepSeek-Coder-V2-Lite  & 181 & 113 & T3 (32\%) & T2 (27\%) \\
\bottomrule
\end{tabular}}
\end{table}

\textbf{(1) False negatives} are the most frequent extraction failure for the
open-source backbones, though GPT-4o inverts the balance
(\cref{tab:errorprofile}).
The extractor misses assumptions that have no identifier-level cue: a T4
assumption encoded as \texttt{users = \{\}} is invisible to any approach that
relies on function names or import statements.
T1 (Input Validation) is a leading source across backbones because its
realizations are often single-line guard checks with no domain-specific
vocabulary.

\textbf{(2) False positives} arise when the extractor proposes an assumption
the code does not implement.
They concentrate in T3 (\cref{tab:errorprofile}), partly reflecting
wrong-category predictions (mode 3 below), which lenient matching counts
twice: as a false positive in the predicted category and a false negative in
the reference category.
A rarer but distinctive pattern is T6 hallucination: the extractor infers
expiry logic or HTTPS enforcement from context, but the code contains
neither, reflecting a tendency in LLMs to over-specify security concerns
when prompted to audit code.
Requiring textual evidence in the code before proposing a T6 assumption is a
straightforward prompt-level mitigation.

\textbf{(3) Wrong-category errors} occur most often at the T1/T3 boundary (guard
clauses), the T2/T4 boundary (return type vs.\ storage), and the T3/T6 boundary
(error handling vs.\ security policy).
These are the same category boundaries that human annotators debated most during
taxonomy development (\S\ref{sec:taxonomy}).
Category errors propagate to downstream dependency mapping, since AST node-type
filters are category-specific.

\begin{figure}[t]
  \centering
  \footnotesize
  \renewcommand{\arraystretch}{1.1}
  \setlength{\tabcolsep}{4pt}
  \begin{tabular}{|p{0.46\columnwidth}|p{0.46\columnwidth}|}
    \hline
    \textbf{(1) False negative} \newline
    \textit{T4: Persistence} \newline
    \texttt{users = \{\}} \newline
    \textit{Missed: ``no I/O; in-memory only.''} &
    \textbf{(2) False positive} \newline
    \textit{T6: Security (hallucinated)} \newline
    \texttt{return token} \newline
    \textit{Hallucinated: ``tokens expire after 1h.''} \\
    \hline
    \textbf{(3) Wrong category} \newline
    \textit{T5 labeled T3} \newline
    \texttt{d = \{\}; d[k] = v} \newline
    \textit{Hash-map choice mis-classified as error policy.} &
    \textbf{(4) Imprecise dependency} \newline
    \textit{T3: full method returned} \newline
    \texttt{def validate(x):} \textit{...entire body...} \newline
    \textit{GT: only \texttt{if x is None: return False}.} \\
    \hline
    \textbf{(5) Failed regeneration} \newline
    \textit{Wrong region modified} \newline
    \texttt{def parse()} \textit{ $\leftarrow$ edited} \newline
    \textit{Should have modified \texttt{def encode()}.} &
    \textbf{(6) Cascading edit} \newline
    \textit{Revising $A_1$ invalidates $A_3$} \newline
    \texttt{bcrypt.hash(pw)} \newline
    \textit{T3 ``return False'' region no longer valid.} \\
    \hline
  \end{tabular}
  \caption{Representative error cases across six failure modes.}
  \label{fig:erroranalysis}
\end{figure}

\textbf{(4) Imprecise dependency} errors occur when the dependency mapper identifies
the correct function but returns its entire body rather than the specific statement.
This is common for T3 assumptions whose realization spans multiple \texttt{return}
statements or \texttt{except} clauses scattered through a function.
The AST refinement pass computes a minimal enclosing sub-tree, which in these
multi-site cases spans the entire function body by construction.
A future fix is to report a \emph{union of minimal sub-trees} rather than a single
enclosing sub-tree when candidates are non-contiguous.

\textbf{(5) Failed regeneration} occurs when the regenerator modifies the wrong
code region.
Root cause: the dependency map for one assumption partially overlaps with a
different function.
For example, a T5 assumption about algorithm choice may be associated with
\texttt{def encode()} via the keyword \texttt{hash}, but the mapper returns
\texttt{def parse()} because it contains more matching tokens.
Disambiguation between candidate regions would require either stronger assumption
descriptions or a secondary prompt to the LLM asking which function best matches.

\textbf{(6) Cascading edits} are the dominant cause of the 17pp syntactic-validity gap
between \sysname and full-regeneration baselines.
When revising assumption $A_1$ (e.g., switching from MD5 to bcrypt), the
regenerator correctly updates the hashing region, but the T3 assumption $A_3$
(``return False on bad password'') was linked to the old hashing pipeline.
After the bcrypt transition, $A_3$'s code region no longer exists in the same
form. Regenerating $A_1$ without updating $A_3$'s region leaves the function
in an internally inconsistent state.
A conflict-detection pass that re-validates all dependency links after each
regeneration step, and flags invalidated regions for secondary review, would
address this without requiring full regeneration.

\subsection{Discussion}
\label{sec:discussion}

The four backbones trade precision against recall in a consistent way.
Qwen2.5-Coder-32B is recall-biased and DeepSeek-Coder-V2-Lite is
precision-biased, which is precisely why combining them in the CW ensemble
yields the best fully cross-model result ($F_1=0.816$).
For code review, recall is usually the more valuable bias: a missed assumption
is costlier than a spurious one a developer can quickly reject.
The trade-off tightens at lower capacity (Qwen2.5-Coder-7B stays precise but
has by far the lowest recall), suggesting that smaller models prefer safe,
high-confidence extractions and that added capacity mainly buys recall.
It nonetheless remains a usable no-API-cost option, comfortably ahead of both
offline baselines.

The AssumptionRecord schema (\cref{lst:schema}) is, structurally, an automated
and fine-grained analogue of an Architectural Decision Record. We develop this
connection, and its implications for architecture knowledge management, in
\S\ref{sec:related}.
A natural extension is to promote \sysname-generated records to first-class ADR
artifacts stored alongside the code, giving a live decision log that grows with
the codebase.

RQ3 exposes a genuine tension between how surgical \sysname is and how reliably
its output compiles.
It makes the smallest edits of any method but reaches only 83\% syntactic
validity, whereas In-Place LLM Edit (IPE) compiles in every case at a larger
edit distance.
What sets \sysname apart is \emph{traceability}: each edit is tied to a specific
AssumptionRecord, so a developer can audit which decision drove which change,
and independent revisions can be applied in parallel, neither of which a
free-form IPE edit supports.
The validity gap stems mainly from cascading edits (error mode~6,
\cref{fig:erroranalysis}), where revising $A_i$ invalidates $A_j$'s region. We
intend to close it with a conflict-detection pass.

In deployment the regeneration step is cheap: across our RQ3 runs it averaged
2.4\,s per task with GPT-4o and 3.6\,s with Qwen32B on local vLLM
(\cref{tab:rq3}), and extraction adds one comparable LLM call.
That overhead is modest against the several seconds generation already takes in
a typical IDE, and the resulting records double as auto-generated design
documentation.
\sysname can be dropped into existing workflows as a post-generation hook in any
IDE extension that wraps an LLM code assistant. Teams adopting it at scale would
accumulate a queryable corpus of assumption records, enabling cross-project
comparison of how different backbones handle recurring ambiguity patterns such
as error-propagation policy (T3) or authentication mechanism (T6).

This paper is a first investigation: it establishes assumption mining as a
problem, contributes a benchmark and taxonomy, and provides an initial empirical
evaluation.
Two deeper questions stay open. First, \emph{why} does an LLM make a particular
assumption (a mechanistic-interpretability question)? Second, can extraction
achieve provably reliable coverage rather than the empirical coverage RQ1
measures? Both need interpretability tooling beyond our current scope.
The self-consistency concern of using LLMs to both generate and extract
assumptions is addressed here through cross-model validation (\S\ref{sec:rq1})
and, going forward, through the human-verification protocol released with the
artifact (\S\ref{sec:threats}).

\section{Related Work}
\label{sec:related}

\subsection{LLM Code Generation, Repair, and Ambiguity}

Large language models achieve strong results on programming benchmarks such as
HumanEval~\cite{chen2021codex}, MBPP~\cite{austin2021mbpp}, and
SWE-bench~\cite{jimenez2024swebench}.
Self-repair approaches~\cite{olausson2023selfrepair,zheng2023selfrefine} improve
correctness by iterating on test failures. Self-planning
methods~\cite{jiang2023selfplanning,dong2023codesurveyllm} decompose prompts
before generation. Conversational tools~\cite{xia2023chatrepair} use test-driven
dialogue to refine incorrect code.
User studies~\cite{vaithilingam2022expectation,ziegler2022productivity} show that
developers frequently struggle to identify \emph{why} generated code is wrong,
a gap that \sysname directly addresses by surfacing implicit design decisions as
editable, traceable records.

Automated Program Repair (APR)~\cite{yuan2022avatar,chen2018sequencer,ye2022rewardrepair}
generates patches for observable failures (test errors, crashes).
\sysname's incremental regenerator (C5) shares the patch-isolation structure,
but its trigger is fundamentally different: APR is \emph{fault-driven}, whereas
C5 is \emph{assumption-driven}, targeting working-but-wrong code that produces
no test failures yet violates developer intent, a regime APR cannot address.

Requirements engineering research addresses natural-language ambiguity
detection~\cite{ferrari2018nlrequire,kiyavitskaya2008ambiguity,ezzini2021nlpre}
in human-written specifications, and a parallel line studies ambiguity in the
\emph{prompts} given to code models: ambiguous, contradictory, or incomplete
task descriptions measurably degrade generated-code
correctness~\cite{larbi2025prompts}, and lightweight classifiers can flag such
defective descriptions before generation~\cite{akli2026defective}.
These approaches all analyze the \emph{input} specification. \sysname instead
extends ambiguity handling to the LLM's \emph{output}, where the artifact under
analysis is executable code and the goal is not only detection but also
structured, targeted revision.

\subsection{Interactive Code Generation and Prompt Engineering}

Several systems augment LLM code generation with interactive clarification steps.
Clarify~\cite{jiang2023selfplanning} generates clarifying questions before
producing code, reducing assumption-driven errors in some settings.
TiCoder~\cite{lahiri2022interactive} formalizes user intent interactively by
generating candidate tests the developer confirms or rejects, pruning the space
of programs consistent with the prompt.
However, these systems require developers to answer questions up-front and do
not link answers to specific code regions, meaning later revision requires a
full reprompt rather than a targeted update.
Prompt engineering surveys~\cite{white2023prompting} document patterns for
eliciting more precise behavior from LLMs, but operate entirely at the prompt
level and cannot recover implicit decisions already made by the model.
\sysname is complementary to both: it operates \emph{post-generation},
diagnosing what the LLM already decided rather than shaping what it will
decide, and enables revision without a full reprompt.

\subsection{Program Analysis and Traceability}

Traceability techniques~\cite{rao2021traceability,mcmillan2011portfolio} link
requirement artifacts to code via information retrieval or machine learning,
and static analysis tools~\cite{livshits2015defense,guo2020graphcodebert}
construct data-flow and control-flow dependency graphs.
Neither class of technique connects dependency edges to the natural-language
intent that produced the code. \sysname bridges this gap with assumption-to-code
dependency graphs for LLM-generated artifacts, enabling developer auditing and
automated change propagation.

Specification mining~\cite{ernst2007daikon} and contract inference extract
program invariants post-hoc from executions or static analysis.
LLM-augmented testing tools (CodaMosa~\cite{lemieux2023codamosa},
Fuzz4All~\cite{xia2024fuzz4all}) leverage language models for coverage-driven
exploration and oracle generation, but all operate \emph{after} code exists
and cannot attribute discovered properties to specific prompt ambiguities.
\sysname is complementary: its assumption layer could seed downstream invariant
checkers, while its primary role is interactive disambiguation \emph{at}
generation time.

\subsection{Design Rationale and Decision Documentation}

Architectural Decision Records (ADRs)~\cite{jansen2005adr,capilla2016adr}
capture the rationale, alternatives, and consequences of design decisions as
lightweight prose documents.
The AssumptionRecord schema in \sysname is directly inspired by this tradition:
both treat design decisions as first-class, addressable artifacts with explicit
alternatives.
The key distinction is \emph{automation and granularity}: ADRs are authored
manually by architects for major architectural choices, whereas \sysname
\emph{generates} fine-grained assumption records automatically during code
synthesis, capturing routine, ephemeral decisions that are too numerous for
manual ADR authoring in practice.
The resulting corpus could retroactively seed architectural documentation
tools~\cite{capilla2016adr}, bridging micro-level assumption tracking and
macro-level architecture knowledge management.

Previous work on automated design-rationale extraction~\cite{yuan2022avatar}
targets source code history and commit messages. \sysname targets the
generation-time decision surface, which is both earlier in the lifecycle and
richer in implicit signals.

\section{Threats to Validity}
\label{sec:threats}

We assess threats following Wohlin et al.~\cite{wohlin2012experimentation}.

\textbf{Internal validity.}
The primary internal threat is that the benchmark reference annotations are
\emph{LLM-derived}: GPT-4o proposed all assumptions and code-region references,
and the taxonomy was developed by a single researcher with LLM assistance.
Hallucinated or mis-categorized reference entries would distort the reported
precision and recall.
To bound this threat, two developers independently reviewed all 111 assumptions
of the curated subset, judging for each (a) whether it is a valid implicit
assumption under Definition~1, (b) its taxonomy category, and (c) whether its
code-region reference is correct (rater background, effort, and completed sheets
in Appendix~\ref{app:benchmark} and the artifact).
Because one of the two annotators is an author, the agreement statistics measure
consistency between the author and a second, independent developer rather than
between two fully external raters. The second developer's independent agreement
nonetheless indicates the labels are not idiosyncratic to the author.
Agreement was high on the two judgments that do not depend on line numbers:
\emph{validity} reached 99.1\% raw agreement ($\kappa=0.66$, and both raters
independently rejected only a single candidate that restates an explicit prompt
requirement), and \emph{category} reached 88.3\% raw agreement ($\kappa=0.85$,
``almost perfect''~\cite{landis1977kappa}).
The $\kappa$ for validity is lower than its raw agreement because $\kappa$ is
deflated when one class dominates (nearly all candidates are valid). We
therefore report both.
The code-region review corroborated the AST-based reference correction of
Appendix~\ref{app:benchmark}: one rater independently flagged the majority of
the automatically corrected references as originally wrong, and after
correction the other rater accepted them.
These results support the reliability of the curated-subset reference.
Residual risk concerns the 150-task extended tier, which is not human-verified
and is used for RQ1 and the secondary RQ2 analysis.
Extraction runs at temperature~0 to minimize non-determinism. Each reported
number comes from a single scored run per task and configuration, with raw
outputs preserved in the artifact.

\emph{LLM backbone bias.}
Because GPT-4o produced the reference annotations, evaluating \sysname
(GPT-4o) on this benchmark risks circular self-consistency rather than true
generalization, which we flag with $\dagger$ throughout
(\cref{tab:rq1main,tab:rq1breakdown}).
We treat the fully cross-model results as our primary validity claims:
Qwen2.5-Coder-32B ($F_1=0.774$) and DeepSeek-Coder-V2-Lite ($F_1=0.771$),
neither of which was involved in benchmark construction, though their scores
still measure agreement with the reference.
The human verification above further reduces the concern that the reference
itself is an artifact of the annotating model.

\emph{Benchmark contamination.}
HumanEval~\cite{chen2021codex} and MBPP~\cite{austin2021mbpp} tasks
(the 150-task extended tier) are part of GPT-4o's training data. The model may
therefore generate predictable, stereotyped solutions whose assumptions are
easier to extract, potentially inflating extraction accuracy relative to novel
tasks.
The 30 tasks of the curated subset are novel and designed to be
out-of-distribution, further reinforcing the primacy of cross-model results on
that subset.

\textbf{External validity.}
The benchmark is Python-only and may not generalize to statically typed languages
(e.g., Rust, Java) where some assumptions are enforced by the type system.
Industrial-scale codebases with multi-file context may expose additional
assumption categories not covered by our six-category taxonomy, and team
conventions or organizational style guides may create assumption patterns not
present in HumanEval/MBPP tasks.

\textbf{Construct validity.}
The headline \AR/\AP/\AFone use category-level matching, which measures
\emph{coverage of decision kinds} rather than description fidelity: two
assumptions in the same category but with different descriptions both count as
correct.
This leniency inflates absolute scores symmetrically across methods, so
relative comparisons remain meaningful but the absolute lenient $F_1$ should
not be read as decision-level accuracy.
To bound the effect we also report the strict decision-level metric
(\S\ref{sec:setup}, \S\ref{sec:rq1}), which requires the predicted description
to match the reference decision. It is substantially lower, quantifying the
gap rather than leaving it unmeasured.
We use embedding cosine (deterministic) rather than an LLM judge for this
metric to avoid compounding the LLM-grades-LLM circularity discussed above.
RQ3's syntactic validity rate is necessary but not sufficient for functional
correctness. Unit-test evaluation is left to future work.

\textbf{Conclusion validity.}
We use a two-tier evaluation design.
RQ1 and RQ3 are evaluated on the full benchmark ($n=180$). RQ2's primary
analysis uses the curated subset ($n=30$ tasks, 109 assumptions), where
code-region references exist, with a secondary full-set analysis.
For $n=180$ comparisons we report bootstrap 95\% confidence intervals
(\S\ref{sec:rq1}, \S\ref{sec:rq3}) and Cliff's~$\delta$ effect sizes
(\S\ref{sec:rq3}). At $n=30$ we omit significance testing and report point
estimates only.
The consistent direction of effects across four independent backbones further
supports conclusion robustness.

\section{Conclusion}
\label{sec:conclusion}

LLM code generation silently embeds implicit assumptions (a mean of 3.8 per
task in our benchmark) that developers have no systematic way to inspect or
revise. These decisions are one source of \emph{working-but-wrong} code: code
that passes available tests yet violates developer intent.

We presented \sysname, a framework that mines implicit assumptions from
LLM-generated code, surfaces them as structured \emph{AssumptionRecord} objects
linked to their governing code regions, and propagates developer-specified
revisions through targeted incremental regeneration.
Offline baselines that rely on surface syntax obtain low scores
($F_1=0.225$ and below), suggesting that such cues alone are insufficient for
this task.
Using only open-source models, \sysname's cross-model ensemble reaches
$F_1=0.816$, a $3.6\times$ improvement over the strongest offline baseline,
and the best result we obtained without a proprietary API.
Under a stricter decision-level metric that also requires the extracted
description to match the reference decision, however, the same ensemble reaches
only $F_1=0.66$ (and the best individual open-source model $0.65$), far above
the near-zero non-LLM baselines but low in absolute terms. Reliably identifying
\emph{which} implicit decision a model made is thus far from solved, and we
hope the benchmark and this second, stricter metric draw future work to the
problem.
Its dependency mapper localizes assumptions to the lines that implement them
more tightly than either baseline, although on binary accuracy it is
competitive with, rather than clearly better than, a keyword heuristic (an
optional LLM-guided variant does better, at extra inference cost).
When a developer revises an assumption, the incremental regenerator changes
$1.3$--$2.4\times$ less code than non-targeted alternatives, but at a cost in
syntactic validity. The error analysis identifies cascading edits as a common
failure mode. Conflict detection is a promising, but untested, mitigation.

The broader implication is that implicit assumptions in AI-generated code are
not merely a quality concern but a \emph{traceability} concern: without explicit
records of why code looks the way it does, developers cannot confidently revise
AI-generated artifacts.
Within the scope of this study, \sysname shows that assumption records can be
generated at interactive-scale cost and used to support auditable code updates.
Whether this workflow improves developer decision making in practice remains
an important question for future user studies.

Future work includes extending human verification to the 150-task extended
tier and beyond, extending \sysname to multi-file repositories, incorporating
project-level style guides as assumption priors, a conflict-detection pass for
cascading edits, description-level quality metrics, unit-test-based evaluation
of regenerated code, and integration with test generation and static analysis
tools.
We release the benchmark and the \sysname implementation (see Data Availability).

\section*{Acknowledgment}
The author thanks Mike Papadakis (University of Luxembourg) for helpful
discussions.

We acknowledge the use of OpenAI Codex and Anthropic's
Claude~\cite{anthropic2025claude} for assistance in polishing the writing of
this paper.

\section*{Data Availability}
The \sysname benchmark (180 tasks / 676 assumptions, 30 curated-subset tasks
for RQ2), the \sysname source code, all evaluation scripts, raw results
for RQ1--RQ3, and the human-verification protocol and annotation sheets, with
instructions to reproduce all tables and figures, are archived on Zenodo at
\url{https://doi.org/10.5281/zenodo.21535058}
(DOI: \texttt{10.5281/zenodo.21535058}).

\bibliographystyle{IEEEtran}
\bibliography{references}

\appendices

\section{Full Extraction Prompt Template}
\label{app:prompt}

The full system and user prompt used by C3 (Assumption Extractor) in all experiments is:

\begin{lstlisting}[language={}, caption={Full extraction prompt.
\texttt{\{prompt\}}, \texttt{\{code\}}, and \texttt{\{schema\}} are substituted at runtime.}]
SYSTEM:
You are an expert software engineer auditing LLM-generated
code for implicit design decisions (implicit assumptions).

USER:
### Original Prompt
{prompt}

### Generated Code
```python
{code}
```

### Task
1. REASON: List every design decision embedded in the code
   that is NOT explicitly required by the prompt. For each,
   name at least one realistic alternative.

2. FORMAT: Return a JSON array of AssumptionRecord objects
   matching this schema:
{schema}

Taxonomy categories:
  T1 -- Input format / validation
  T2 -- Return type / output structure
  T3 -- Edge-case / error-handling policy
  T4 -- Persistence / storage backend
  T5 -- Algorithm / performance trade-off
  T6 -- Security / authentication policy

Return ONLY valid JSON after the tag [RECORDS].
Do not include any other text after [RECORDS].

[RECORDS]
\end{lstlisting}

The schema object inserted at \texttt{\{schema\}} is:

\begin{lstlisting}[language={}]
{
  "id":           "string (e.g. 'A1')",
  "category":     "T1|T2|T3|T4|T5|T6",
  "description":  "natural-language statement of the assumption",
  "rationale":    "why the LLM made this choice",
  "alternatives": ["list of >=1 realistic alternatives"],
  "confidence":   "float in [0,1]",
  "severity":     "low|medium|high"
}
\end{lstlisting}

The \texttt{code\_refs} field (used by C4) is not included in the extraction
prompt. C4 populates it separately after the extraction call returns.

\section{Full Per-Category Precision, Recall, and $F_1$}
\label{app:tables}

\cref{tab:rq1full} extends \cref{tab:rq1breakdown} with per-category \AR and
\AP values for the cross-model backbones.

\begin{table*}[htbp]
\caption{Full per-category precision (\AP), recall (\AR), and $F_1$ (\AFone)
         for all \sysname backbones on the benchmark ($n=180$, 676 assumptions).
         $\dagger$ self-evaluation risk for GPT-4o on 150/180 tasks.}
\label{tab:rq1full}
\resizebox{\textwidth}{!}{%
\begin{tabular}{@{}l l rrr rrr rrr rrr rrr rrr@{}}
\toprule
& & \multicolumn{3}{c}{\textbf{T1}} & \multicolumn{3}{c}{\textbf{T2}} & \multicolumn{3}{c}{\textbf{T3}} & \multicolumn{3}{c}{\textbf{T4}} & \multicolumn{3}{c}{\textbf{T5}} & \multicolumn{3}{c}{\textbf{T6}} \\
\cmidrule(lr){3-5}\cmidrule(lr){6-8}\cmidrule(lr){9-11}\cmidrule(lr){12-14}\cmidrule(lr){15-17}\cmidrule(lr){18-20}
\textbf{Method} & \textbf{All $F_1$} & P & R & $F_1$ & P & R & $F_1$ & P & R & $F_1$ & P & R & $F_1$ & P & R & $F_1$ & P & R & $F_1$ \\
\midrule
GPT-4o$^\dagger$ & 0.884
  & 0.876 & 0.936 & 0.905
  & 0.930 & 0.839 & 0.882
  & 0.823 & 0.922 & 0.869
  & 0.800 & 0.500 & 0.615
  & 0.866 & 0.940 & 0.902
  & 0.429 & 0.600 & 0.500 \\
Qwen2.5-Coder-32B & 0.774
  & 0.865 & 0.601 & 0.709
  & 0.934 & 0.692 & 0.795
  & 0.679 & 0.892 & 0.771
  & 0.600 & 0.750 & 0.667
  & 0.834 & 0.868 & 0.851
  & 0.429 & 0.600 & 0.500 \\
DeepSeek-Coder-V2-Lite & 0.771
  & 0.876 & 0.734 & 0.799
  & 0.788 & 0.804 & 0.796
  & 0.818 & 0.651 & 0.725
  & 0.556 & 0.625 & 0.588
  & 0.808 & 0.755 & 0.781
  & 0.400 & 0.800 & 0.533 \\
Qwen2.5-Coder-7B & 0.593
  & 0.923 & 0.414 & 0.571
  & 0.738 & 0.867 & 0.797
  & 0.904 & 0.283 & 0.431
  & 0.375 & 0.375 & 0.375
  & 0.824 & 0.371 & 0.511
  & 0.667 & 0.400 & 0.500 \\
\bottomrule
\end{tabular}}
\end{table*}

\section{Benchmark Construction Details}
\label{app:benchmark}

\subsection{Novel Task Design}
The 30 novel tasks of the curated subset were designed to:
(a) cover domains underrepresented in standard benchmarks (authentication,
serialization, file I/O, network I/O, concurrency);
(b) stress-test T4 (Persistence) and T6 (Security) categories by requiring
tasks where storage backend and security policy choices are meaningful;
(c) be short enough that GPT-4o could solve them in under 30 lines of Python,
avoiding multi-file complexity; and
(d) be genuinely ambiguous: each task deliberately omits at least two
meaningful design choices.

\subsection{Reference Annotation Pipeline}
For every task, GPT-4o~\cite{openai2024gpt4o} extracted an AssumptionRecord list using the standard
prompt (\cref{app:prompt}) at temperature~0. Records were then validated
programmatically (schema conformance, category vocabulary, deduplication by
category and description similarity).
For the curated subset, code-region references (line ranges) were produced by
the same pipeline from the \texttt{code\_refs} field.
Because these line numbers are LLM-counted rather than parser-derived, they
are susceptible to indexing errors: a tree-sitter consistency scan (each
reference's declared AST-node type checked against the node actually present
at the stored lines) found 15 of the 111 curated references off by one to two
lines, consistent with sporadic 0-indexed counting by the annotating model.
These were corrected by snapping each reference to its declared node, with
every correction manually adjudicated against the assumption description
(\texttt{scripts/fix\_coderef\_offsets.py} in the artifact documents all 15). All RQ2 results in this paper are scored against the corrected references.

\emph{Human verification.}
Two developers then independently reviewed all 111 curated-subset assumptions
using the released protocol, recording for each a validity judgment, a taxonomy
category, and whether the (corrected) code-region reference is correct.
Both have at least five years of professional software-development experience in
industry. One is an author of this paper and the other is an independent
developer not otherwise involved in the project.
Each rater spent roughly four hours on this independent pass, after which the two
met for an approximately one-hour joint session to resolve every disagreement to
consensus.
Inter-rater agreement (before reconciliation) was 99.1\% ($\kappa=0.66$) on
validity and 88.3\% ($\kappa=0.85$) on category. On the code-region judgment the
raters agreed on 89.6\% of comparable references.
The region review also served as an independent audit of the AST correction
above: one rater, working from the pre-correction line numbers, flagged the
majority of the 15 corrected references as originally wrong, while the other,
working from the corrected numbers, accepted them, an independent
confirmation of the fix.
Disagreements were resolved by discussion to form the final reference: two
assumptions were removed as invalid, eight category labels were corrected, and
twelve code-region spans were reconciled (several additional off-by-one
references the automated scan had missed), yielding a human-verified curated
subset of 109 assumptions.
All curated-subset (RQ2) and open-source RQ1 results in this paper are scored
against this reconciled reference. Re-scoring changed the full-benchmark
numbers by at most 0.006, confirming the LLM-derived reference was already
reliable.
The rating sheets, the agreement script, the adjudication decisions, and the
resulting \texttt{consensus.jsonl} are included in the artifact.

\subsection{Extended-Tier Tasks}
The 150 tasks drawn from HumanEval and MBPP use the same pipeline.
These annotations are used only for RQ1 extraction evaluation and the RQ2
full-set secondary analysis (\cref{tab:rq2}, auto-labeled columns). RQ2's
primary analysis uses only the curated subset.
The extended tier is not human-verified. This is noted as a residual threat in
\S\ref{sec:threats}.

\section{Conf-Weighted Ensemble Algorithm}
\label{app:ensemble}

Algorithm~\ref{alg:ensemble} gives the pseudocode for the Conf-Weighted (CW)
ensemble used to combine the high-recall backbone (Qwen2.5-Coder-32B) and
high-precision backbone (DeepSeek-Coder-V2-Lite).

\begin{figure}[htbp]
\begin{lstlisting}[language=Python, caption={Conf-Weighted (CW) ensemble
  (as implemented in \texttt{scripts/run\_rq1\_ensemble.py}).
  \texttt{preds\_a} and \texttt{preds\_b} are AssumptionRecord lists from the
  two backbones for one task.},
  label={alg:ensemble}]
def ensemble_conf_weighted(preds_a, preds_b,
                           threshold=0.5):
    """Include a category if the mean
    confidence across all instances from
    either backbone >= threshold; keep the
    highest-confidence instance."""
    by_cat = defaultdict(list)
    for p in preds_a + preds_b:
        by_cat[p["category"]].append(p)

    result = []
    for cat, items in sorted(by_cat.items()):
        mean_conf = (sum(x["confidence"]
                     for x in items)
                     / len(items))
        if mean_conf >= threshold:
            best = max(items,
                key=lambda x: x["confidence"])
            result.append(best)
    return result
\end{lstlisting}
\end{figure}

Because both backbones vote on each category and only the single
highest-confidence instance per qualifying category is kept, CW trades a small
amount of recall for precision relative to a plain union.
On the benchmark the three strategies score:
\textbf{Union} (keep all records from either backbone) $P=0.755$, $R=0.857$,
$F_1=0.803$;
\textbf{Intersection} (keep only categories produced by both backbones)
$P=0.887$, $R=0.624$, $F_1=0.733$;
\textbf{CW} $P=0.810$, $R=0.823$, $F_1=0.816$.
CW outperforms both by combining the backbones' confidence signals rather than
applying a binary gate.

\section{Software Artifact Details}
\label{app:software}

\sysname is implemented as an open-source Python package
(\texttt{assumption\_miner}) released alongside this paper.
This section describes the software architecture, interfaces, and replication support.

\renewcommand{\dbltopfraction}{0.9}
\renewcommand{\textfraction}{0.08}
\renewcommand{\dblfloatpagefraction}{0.7}
\begin{figure*}[!t]
  \centering
  \setlength{\fboxsep}{0pt}\setlength{\fboxrule}{0.4pt}%
  \fbox{\includegraphics[width=0.97\textwidth]{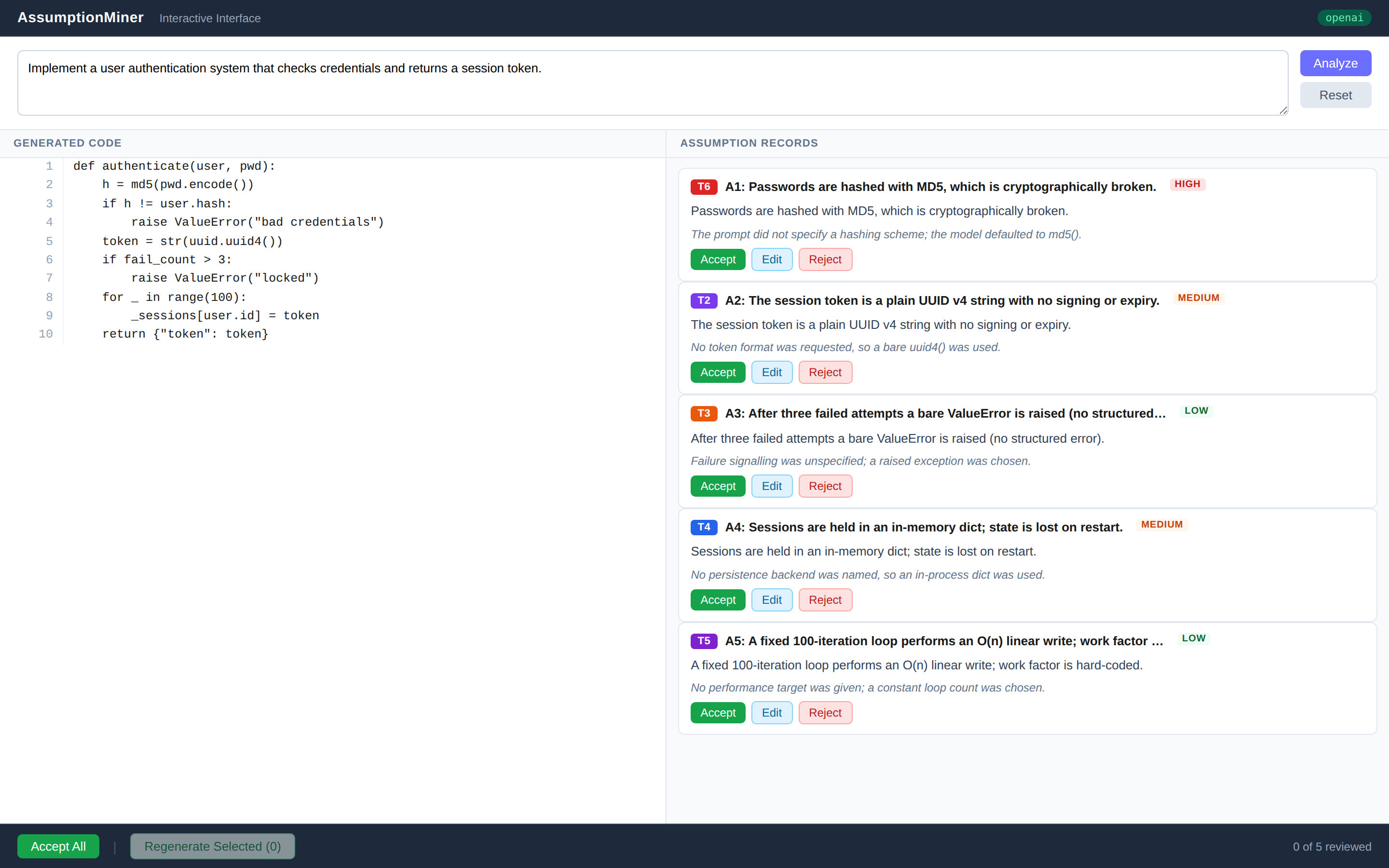}}\\[4pt]
  \fbox{\includegraphics[width=0.97\textwidth]{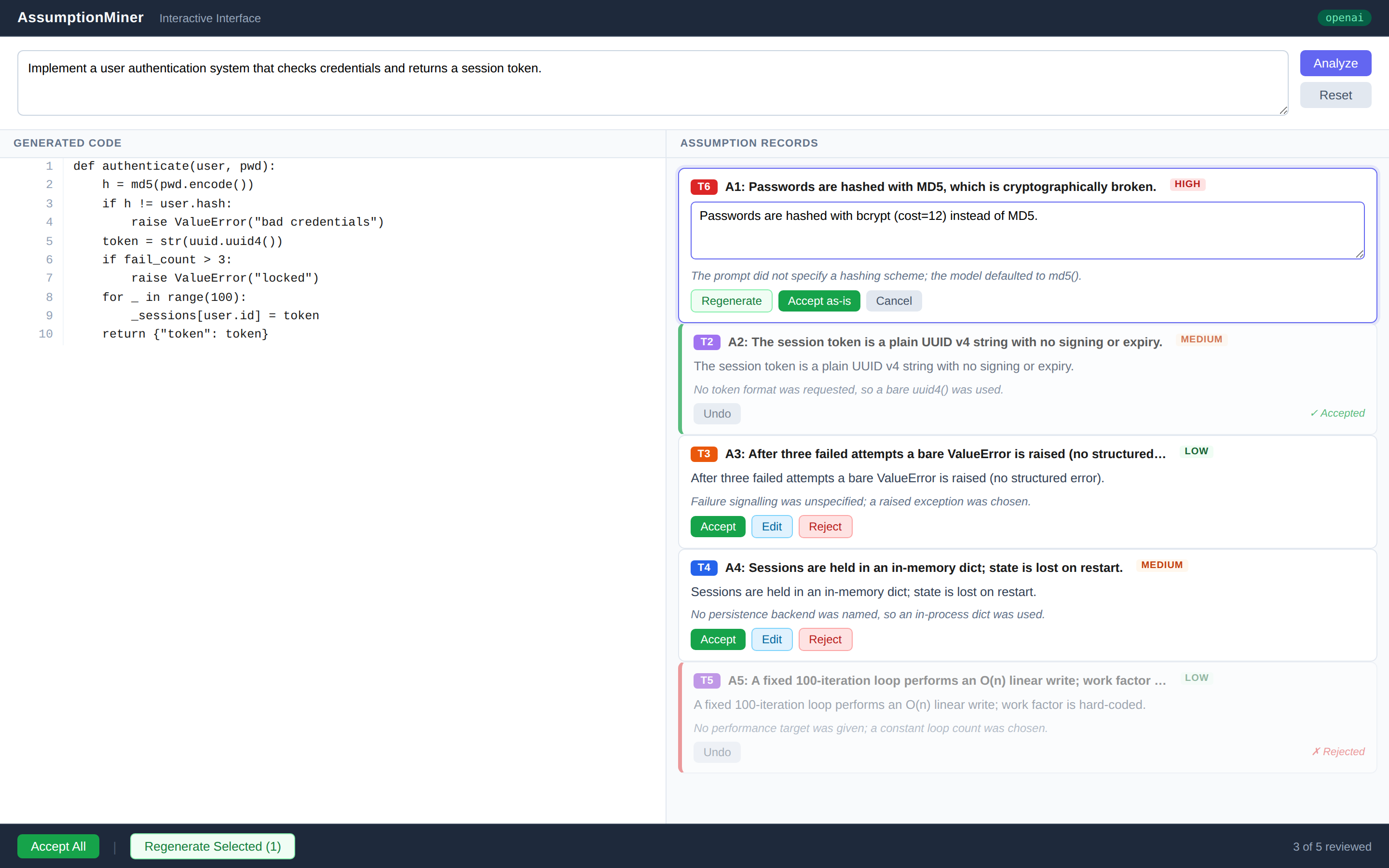}}
  \caption{The \sysname web interface on the running authentication example
           (screenshots of the released \texttt{interface/} application).
           \textbf{Top:} after \emph{Analyze}, the backbone-generated code (left)
           is shown beside its five extracted assumption records (right), each
           labeled with its taxonomy category (T2--T6) and an LLM-assigned
           severity. Hovering a card highlights the code lines it governs.
           \textbf{Bottom:} mid-review, the developer has accepted A2, rejected
           A5, and opened A1 for editing (MD5\,$\to$\,bcrypt). Pressing
           \emph{Regenerate} invokes the incremental regenerator (C5) on only
           that assumption's code region. The footer tracks review progress.}
  \label{fig:interface_real}
\end{figure*}

\subsection{Architecture and Components}

The package maps directly to the five components described in
\cref{sec:architecture}, one module per component:

\textbf{C1 (Prompt Ingestion), \texttt{ingestion.py}.} A \texttt{CodeSample}
data class with three constructors: \texttt{from\_strings} (prompt and code as
plain strings), \texttt{from\_json\_file} (iterates benchmark JSON files such
as \texttt{data/benchmark.json}), and \texttt{from\_code\_file} (prompt plus a
path to a source file).

\textbf{C2 (Code Generator), \texttt{generator.py}.} \texttt{generate\_code()}
dispatches to interchangeable backends selected by the
\texttt{ASSUMPTION\_MINER\_BACKEND} environment variable: \texttt{openai}
(OpenAI API, configurable model) or \texttt{local} (an OpenAI-compatible vLLM
endpoint for open-source models).

\textbf{C3 (Assumption Extractor), \texttt{extractor.py}.}
\texttt{extract\_assumptions()} issues the structured two-part prompt
(Appendix~\ref{app:prompt}) and parses the \texttt{[RECORDS]} JSON block into
\texttt{AssumptionRecord} objects (\texttt{schema.py}, which also defines the
\texttt{CodeRef} type for dependency links).

\textbf{C4 (Dependency Mapper), \texttt{dependency.py}.}
\texttt{map\_dependencies()} implements the two-pass strategy:
keyword-overlap candidate selection followed by AST narrowing that snaps
candidate lines to the smallest enclosing category-appropriate node.
The LLM-guided variant evaluated in \S\ref{sec:rq2ablation} is available via a
\texttt{use\_llm} flag. The AST heuristic is the default for its zero
inference cost, with the LLM-guided mapper available for accuracy-critical
settings.

\textbf{C5 (Incremental Regenerator), \texttt{regenerator.py}.}
\texttt{regenerate()} accepts a revised \texttt{AssumptionRecord} and the
original code, constructs a region-targeted prompt from the dependency map
($k=5$ context lines), and splices the regenerated region back into the
surrounding code.

The offline baselines of \cref{tab:rq1main} (CE, KBE) and the prompting
baselines (DG, CQ, CoT) are implemented in \texttt{baselines.py}, so every
number in \cref{sec:evaluation} is reproducible from the same package.

\subsection{Scripts and Entry Points}

All experiment scripts are under \texttt{scripts/}:

\begin{itemize}[noitemsep]
  \item \texttt{run\_rq1.py}: extraction evaluation on a benchmark file.
  \item \texttt{run\_rq1\_offline.py}: CE/KBE offline baselines.
  \item \texttt{run\_rq1\_ensemble.py}: Union/Intersection/CW ensembles
        combining two extractor outputs.
  \item \texttt{run\_rq1\_semantic.py}: strict decision-level \AFone{} via
        sentence-embedding similarity at threshold $\tau$.
  \item \texttt{build\_strict\_inputs.py}: reconstructs per-prediction
        descriptions for the CE/KBE baselines and the CW ensemble so they can
        be scored under the strict metric.
  \item \texttt{run\_rq2.py}, \texttt{run\_rq2\_ablation.py},
        \texttt{run\_rq2\_usellm.py}: dependency accuracy, node-type
        ablation, and the heuristic-vs.-LLM mapper comparison.
  \item \texttt{run\_rq3.py}: code adaptability evaluation.
\end{itemize}

Each script accepts \texttt{--benchmark}, \texttt{--out}, \texttt{--backend},
and \texttt{--limit} flags documented in \texttt{--help} output. A
\texttt{--dry-run} flag substitutes the offline KBE extractor for all LLM
calls, enabling smoke-testing without API credentials or GPU access.

\subsection{Interactive Interface}

The interactive interface (\S\ref{sec:interface}) is a single-page web
application served by \texttt{interface/server.py}, exposing three REST
endpoints: \texttt{/api/analyze} (prompt $\to$ code + assumption records),
\texttt{/api/regenerate} (code + revised assumption $\to$ updated code), and
\texttt{/api/status}.
It runs against either backend (OpenAI or local vLLM) using the same
environment-variable configuration as the package.
\cref{fig:interface_real} shows the running interface on the paper's
authentication example: the generated code appears on the left, and each
extracted assumption becomes a category-tagged, severity-scored card on the
right that the developer can accept, edit, or reject before regenerating.

\subsection{Extensibility}

New backbone LLMs are added as a dispatch case in
\texttt{generator.py}/\texttt{extractor.py} (any OpenAI-compatible endpoint
works out of the box via the \texttt{local} backend).
New assumption categories require extending the taxonomy vocabulary in the
extraction prompt and the category-to-node-type map in \texttt{dependency.py}.

\subsection{Quality and Testing}

The package ships with a unit-test suite of 119 tests across eight modules
(\texttt{assumption\_miner/tests/}), all passing at release, covering schema serialization round-trips, dependency-mapper
AST logic, extractor output parsing, ingestion, the offline baselines, and
the RQ evaluation helpers.
All results in \cref{sec:evaluation} were generated from the released code
with no post-hoc modifications.

\section{Reproducibility Notes}
\label{app:repro}

\subsection{Environment}
All local-model experiments were conducted on a machine with an NVIDIA RTX PRO
6000 Blackwell Max-Q Workstation Edition GPU (96\,GB VRAM), serving models
through vLLM's OpenAI-compatible server.
The package requires Python~$\geq$3.10 with \texttt{tree-sitter}~$\geq$0.22 and
\texttt{tree-sitter-python}~$\geq$0.22 (exact dependency pins are recorded in
\texttt{pyproject.toml} in the artifact).
Cloud API calls used the OpenAI API with model alias \texttt{gpt-4o}.

\subsection{Running the Experiments}
From the repository root:

\begin{lstlisting}[language=bash, numbers=none]
# Install dependencies
pip install -e ".[dev]"

# RQ1 - GPT-4o (cloud)
python scripts/run_rq1.py \
  --backend openai --model gpt-4o \
  --benchmark data/benchmark.json \
  --out data/results/rq1_gpt4o.json \
  --delay 0.5

# RQ1 - Qwen2.5-Coder-32B (local vLLM)
bash scripts/start_vllm_server.sh  # terminal 1
python scripts/run_rq1.py \
  --backend local \
  --benchmark data/benchmark.json \
  --out data/results/rq1_qwen32b.json \
  --delay 0.1  # terminal 2

# RQ1 ensemble (no GPU if using cloud backbones)
python scripts/run_rq1_ensemble.py \
  --benchmark data/benchmark.json \
  --out data/results/rq1_ensemble.json

# RQ2 mapper ablation
python scripts/run_rq2_usellm.py \
  --backend openai --model gpt-4o \
  --benchmark data/benchmark.json \
  --out data/results/rq2_use_llm_gpt4o.json

# RQ3 code adaptability
python scripts/run_rq3.py \
  --backend openai --model gpt-4o \
  --benchmark data/benchmark_mini.json \
  --out data/results/rq3_gpt4o.json
\end{lstlisting}

\noindent A dry-run mode (no API calls) is available for smoke testing:

\begin{lstlisting}[language=bash, numbers=none]
python scripts/run_rq1.py --dry-run \
  --benchmark data/benchmark_mini.json --limit 3
\end{lstlisting}

\subsection{Randomness}
LLM outputs are non-deterministic in general. We minimize this at the source.
Assumption extraction (C3) runs at \texttt{temperature=0}. Code generation
(C2) and regeneration (C5) run at \texttt{temperature=0.2}
(RQ3 baselines likewise at 0.2).
Each reported metric comes from a single scored run per task and
configuration. Raw model outputs for every run are preserved in the artifact,
and the bootstrap confidence intervals in \S\ref{sec:rq1} and \S\ref{sec:rq3}
quantify across-task variability.

\begin{IEEEbiography}[{\includegraphics[width=1in,height=1.25in]{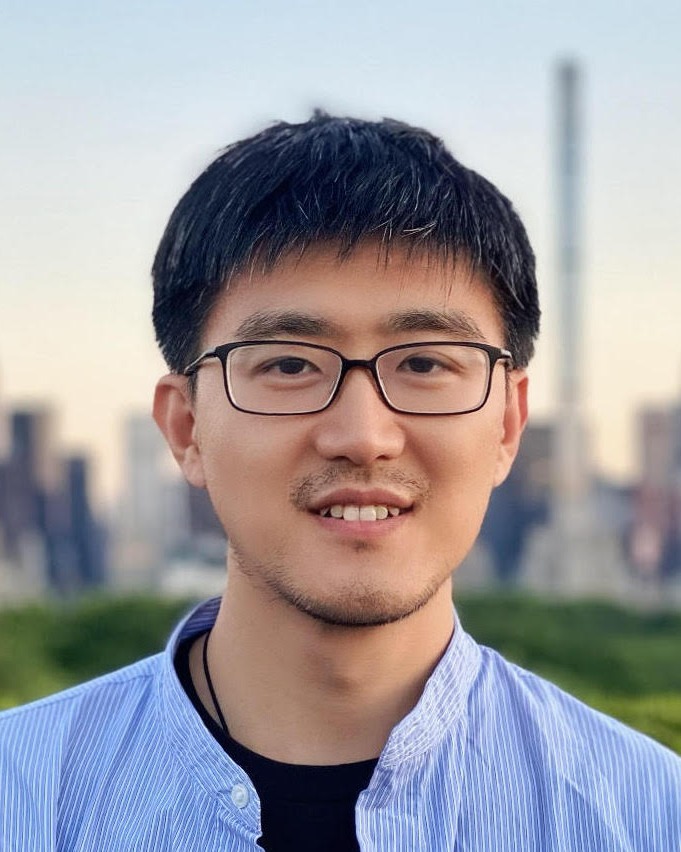}}]{Jie ``JW'' Wu}
is an Assistant Professor in the Department of Computer Science at Michigan
Technological University, USA, starting Fall 2025.
He was a Postdoctoral Fellow at the University of British Columbia Okanagan,
Canada, from 2024 to 2025.
He worked as a full-time software developer for eight years at Snap Inc.\ and
Microsoft in USA.
His recent research centers on intent-aware software engineering,
developing AI-assisted techniques that infer, communicate, and preserve
developer intent to enable the development of high-quality software.
\end{IEEEbiography}
\vspace*{\stretch{40}}

\end{document}